\documentclass[a4paper,fleqn,usenatbib]{mnras}

\usepackage{newtxtext,newtxmath}

\usepackage[T1]{fontenc}
\usepackage{ae,aecompl}

\usepackage{graphicx}	
\usepackage{amsmath}	
\usepackage{amssymb}	
\usepackage{color}
\usepackage{xspace}
\usepackage{enumitem}

\hypersetup{draft}

\newcommand{\loverm}{L$_{bol}$/M$_{clump}$}

\newcommand{\hii}{H{\sc ii}}
\newcommand{\mcl}{M$_{CL}$}
\newcommand{\cht}{CH$_3$C$_2$H(J=12-11)}
\newcommand{\ch}{CH$_3$C$_2$H}
\newcommand{\msun}{M$_{\odot}$}
\newcommand{\lsun}{L$_{\odot}$}

\newcommand{\msunyr}{M$_{\odot}$\,yr$^{-1}$}

\newcommand{\um}{$\mu$m}                                 

\newcommand{\higal}{Hi-GAL}
\newcommand{\fcore}{$f_{\rm core}$\xspace}
\newcommand{\mmean}{$\overline{M}_{\star}$}

\newcommand{\gsim}{\;\lower.6ex\hbox{$\sim$}\kern-7.75pt\raise.65ex\hbox{$>$}\;}
\newcommand{\gl}{\;\lower.6ex\hbox{$<$}\kern-7.75pt\raise.65ex\hbox {$>$}\;}
\newcommand{\lsim}{\;\lower.6ex\hbox{$\sim$}\kern-7.75pt\raise.65ex\hbox{$<$}\;}

\newcommand{\mclump}{$M_{\rm clump}$\xspace}
\newcommand{\sftime}{$t_{\rm SF}$\xspace}
\newcommand{\tdust}{$T_{\rm dust}$\xspace}
\newcommand{\Myr}{Myr\xspace}
\newcommand{\tstar}{$t_{\star}$\xspace}
\newcommand{\mstar}{$M_{\star}$\xspace}
\newcommand{\microns}{$\mu$m\xspace}

\title[A new grid of protoclusters SED models]{Evolution of young protoclusters embedded in dense massive clumps. A new grid of population synthesis SED models and a new set of L/M evolutionary tracks.}

\author[S. Molinari et al.]{Sergio Molinari,$^{1}$\thanks{E-mail: molinari@iaps.inaf.it}
Adriano Baldeschi,$^{1}$
Thomas P. Robitaille,$^{2}$
Esteban F. E. Morales,$^{2, 5}$
\newauthor{Eugenio Schisano,$^{1}$
Alessio Traficante,$^{1}$
Manuel Merello,$^{1}$
Marco Molinaro,$^{3}$}
\newauthor{Fabio Vitello,$^{4}$
Eva Sciacca,$^{4}$
and Scige J. Liu$^{1}$}
\\
$^{1}$Istituto Nazionale di Astrofisica - IAPS, Via Fosso del Cavaliere 100, I-00133 Roma, Italy\\
$^{2}$Max Planck Institute for Astronomy, K\"onigstuhl 17, D-69117 Heidelberg, Germany\\
$^{3}$Istituto Nazionale di Astrofisica - Osservatorio Astronomico di Trieste, Via Bazzoni 2, I-34124 Trieste, Italy\\
$^{4}$Istituto Nazionale di Astrofisica - Osservatorio Astrofisico di Catania, Via S. Sofia 78, I-95123 Catania, Italy\\
$^{5}$I. Physikalisches Institut, Universit\"at zu K\"oln, Z\"ulpicher Str. 77, 50937, K\"oln, Germany.\\
}

\date{Accepted XXX. Received YYY; in original form ZZZ}

\pubyear{2017}

\begin{document}
\label{firstpage}
\pagerange{\pageref{firstpage}--\pageref{lastpage}}
\maketitle

\begin{abstract}
A grid of 20 millions 3-1100\um\  SED models is presented for synthetic young clusters embedded in dense clumps. The models depend on four primary parameters: the clump mass \mclump\ and dust temperature \tdust, the fraction of mass \fcore\ locked in dense cores, and the age of the clump \sftime. We populate the YSO clusters  using the \cite{Kroupa2001} IMF and the YSOs SED models grid of \cite{rob06}.

We conduct extensive testing of SED fitting using a simulated dataset and we find that \mclump\ essentially depends on the submillimeter portion of the SED, while \tdust\ is mostly determined from the shape of the SED in the 70-350\um\ range. Thanks to the large number of models computed we verify that the combined analysis of L/M, [8-24] and [24-70] colours removes much of the SEDs \fcore -\sftime\ degeneracy. 

The L/M values are particularly useful to diagnose \fcore. L/M$\leq$1 identifies protoclusters with \fcore$\leq$0.1 and \sftime $\lsim 10^5$ years, while L/M$\gsim$10 excludes \fcore$\leq$0.1. We characterize lower limits of L/M where ZAMS stars are not found in models, and we also find models with L/M $\geq$10 and no ZAMS stars, in which [8-24]$\gsim0.8\pm 0.1$ independently from \mclump, temperature and luminosity.

This is the first set of synthesis SED models suited to model for embedded and unresolved clusters of YSOs. A set of new evolutionary tracks in the L/M diagram is also presented. 
\end{abstract}

\begin{keywords}
stars: formation -- stars: protostars -- ISM:general
\end{keywords}



\section{Introduction}

High mass stars in the Galaxy are formed in very dense and compact structures in clouds called ``clumps'' (forming star clusters), and structures inside them called ``cores'', forming single stars or small multiple systems (e.g., \citealt{BT2007}). The energy radiated by Young Stellar Objects (YSOs) is generally produced by accretion of surrounding material onto the protostellar core from a circumstellar disk and/or a hosting envelope that embeds it, and by the contracting protostellar core itself \citep{PS1993}. Such energy, radiated close to the forming star, is absorbed and re-emitted by dust along its path outward and escapes the system as low-energy radiation. Depending on the relative importance of the protostellar core as opposed to a circumstellar disk or envelope, the geometry of the system with respect to the observer, and the evolutionary stage of the YSO, the shape of the emerging Spectral Energy Distribution (SED) shows a large diversity from the near-infrared to the millimeter spectral range \citep{rob06}. 

While YSOs in nearby star-forming regions can be resolved and modelled as individual sources across the electromagnetic spectrum, star forming regions located more than a few hundred pc away are more complex to analyze, due to the blending of multiple sources into a single resolution element as we go from the near and mid-infrared to the far-IR and submm where instrumental diffraction-limited beams are larger. To model single YSO sources, \cite{rob06} (hereafter R06) presented a grid of radiative transfer models of young stellar objects, which ranged from models of protostars to models of pre-main-sequence stars surrounded by protoplanetary disks, for a wide range of stellar masses. These models are extensively used to infer YSOs and circumstellar parameters from broad band photometry (e.g., \citealt{elia10, Yadav+2016}). However, stars mostly form in clusters and the issue of blending, where single objects at distances of a few kpc are in reality compact embedded clusters, becomes important. Indeed, the estimated size of compact far-IR clumps in the Hi-GAL survey ranges from 0.1 to 1pc \citep{Elia+2017}, that is one order of magnitude above the typical size of envelopes hosting single or binary forming YSOs. Such objects are most likely hosting forming clusters of YSOs, something that becomes more apparent when such objects are observed at higher spatial resolution in the near/mid-IR and in the millimeter (e.g., \citealt{Yun+2015, Beuther+2015, Heyer+2016}).

The process of associating counterparts at different wavelengths with a pure positional match, normally termed band-merging, can be quite complex if we would like to obtain the maximum wavelength coverage for the SEDs, going from the submm and far-IR like Hi-GAL , ATLASGAL and BGPS (\citealt{Molinari+2016a, Schuller2009, Aguirre2011}) to the mid and near-IR like GLIMPSE, WISE and MIPSGAL (\citealt{Benjamin+2003, Wright+2010, Carey+2009}), with the additional complication that the flux contribution of individual objects in confused bands will also be function of the evolutionary stage of each object in the clump \citep{Baldeschi+2017a}. Modelling these sources correctly is crucial, since estimates of the star formation rate and efficiency in Galactic star formation regions often rely on the implicit assumption that the objects are single. While this is a good approximation for typical (massive) YSOs detected by GLIMPSE in the mid-infrared with $\sim$2\arcsec\ resolution \citep{Morales+2017}, it is unlikely to be the case for most compact sources revealed in the far-IR or submillimeter by single-dish facilities (e.g., Herschel) on the Galactic Plane (see above). In the case that multiple objects are indeed present in observed far-IR compact sources, assuming that sources are single objects would result in different star formation rates and efficiencies than would have been derived if higher-resolution data had been available and the multiple sources had been resolved \citep{Veneziani+2017}. This issue is particularly pressing for the systematic  analysis of extensive compact source catalogues produced by the latest generation of Galactic Plane surveys in the infrared and submillimeter \citep{Molinari+2016a, Elia+2017, Ginsburg2013, Csengeri+2014}, where the objects that are detected are in their vast majority clumps large enough to host multiple YSOs, and not cores.

Constructing protocluster SEDs should ideally be done by taking a realistic density and source distribution, for example from a simulation of a forming cluster, and carrying out full 3D radiative transfer calculations to determine the temperature at every point, and from this the overall SED. However, doing this over a wide dynamic range of cluster sizes and evolutionary stages is computationally prohibitive. Therefore, the approach we take in this paper is to approximate protocluster SEDs as being the sum of the SEDs of individual young stars, as well as a contribution to the long-wavelength SED from cold dust in the intra-clump material. The aim is indeed not to provide the most realistic individual models, but rather to produce a large number of models that can be use to explore broad trends over parameter space. It is important to remember throughout the paper that some combinations of parameters may produce unrealistic systems like, e.g., large clusters of luminous YSOs and very low intra-clump dust temperature, or systems with efficiency 1 in going from cores to stars. Using this models grid to fit observed SEDs requires that the physical plausibility of the best-fit models is always checked.

We will describe the assumptions and the methods to generate the models SEDs in section \ref{modelsgrid}. Section \ref{analysis} will present a full analysis of the parameters space of SED models and show how the shape of the model  SEDs depend on the choices of the input parameters. Section \ref{tracks} presents a new set of evolutionary tracks for cluster-hosting clumps in the  L/M diagram, that supersedes the evolutionary tracks of \cite{Molinari+2008}. Finally Section \ref{applications} will show the application of the models grid to the fitting of observed SEDs for two massive clumps in the \higal\ survey, while conclusions will be summarised in Section \ref{conclusions}.

\section{A grid of synthetic protocluster models}

\label{modelsgrid}

Each protocluster is defined as a clump, a fraction of which is in the form of compact sources/protostars, and the rest is in the form of diffuse intra-cluster gas/dust. The models are described by the following parameters:

\begin{itemize}[leftmargin=*]

\item The total clump mass \mclump, which is taken to be a time-invarient quantity, including the mass of protostars (forming YSOs + disk + envelope), any ZAMS stars that may have formed, and diffuse gas/dust

\item The fraction \fcore of the mass in compact cores (forming stars)

\item The time \sftime since star formation began in the clump 

\item The diffuse dust temperature \tdust\ in the clump

\end{itemize}

Rather than set up a regular grid of models, we take the approach of randomly sampling \mclump, \fcore, and \tdust\ within pre-defined ranges. We then compute the models for each set of these three parameters for a number of values of \sftime, so that we can follow the evolution of a given model over time (in that respect, the models are sampled in a grid only along the \sftime parameter).

The clump mass \mclump was randomly sampled logarithmically between $10$ and $10^5$\,\msun to cover the full range of compact clump masses found by \cite{Elia+2017} on the Galactic plane with Herschel. No models were computed for clump masses below 10\msun\ as these are not likely to produce intermediate mass stars (M$\geq$3\msun) for plausible star formation efficiencies (see below). 

The fraction of mass in compact sources \fcore was randomly sampled linearly between 0 and 50\%, to cover from the youngest clumps, where no or little stars have yet formed, to more evolved clumps in which both core and star formation may have taken place for longer times. 

The time \sftime since star formation has begun in the core was regularly and logarithmically sampled using 20 values between 0.01 and 0.5\,\Myr, and we assume that the stars have formed at a constant star formation rate over this time. Young clusters that are revealed in the near-IR and at optical wavelengths have dispersed most of their clump natal material and can have a dense core/clump material fraction higher than 50\%, but they typically have ages of a few to several million years. Our choice of sampling YSOs with ages up to 5$\times 10^5$ years qualitatively justifies the adopted limit on \fcore\ (see above).

Finally, the dust temperature \tdust for the diffuse dust was randomly sampled between 10 and 30\,K, representative of the average dust temperatures in dense clumps from Herschel \citep{Elia+2017}, independently with respect to the other model parameters. This choice might in principle lead to unphysical models and we will discuss this issue in more detail in \S\ref{balance}.

In total, we sample $10^5$ combinations of \mclump, \fcore, and \tdust, and we then compute models for 20 different ages for each of these, giving a total of $2\times10^6$ sets of parameters. For each of these parameters sets, we set up protocluster models in the following way. We started off by computing the total mass in compact sources based on  \mclump\ and \fcore . We then randomly sampled a mass function between 0.1\msun\ and 50\msun, until the mass reached the desired total compact mass. We assumed that the mass of the compact sources (including protostar, disk and envelope) follows an initial mass function given by \cite{Kroupa2001}. Since the randomly sampled masses never add up to exactly the desired total compact mass, we stopped half the time when the mass was just below versus just above the desired compact source mass. For each compact source, we then random sampled uniformly an age between 0 and \sftime.

With the compact source masses and ages in hand, we used the set of model SEDs by R06 to determine the observable properties of each source. The models in R06 are model SEDs for individual young stellar objects that include circumstellar disks and envelopes. Those models did not cover parameter space uniformly -- instead, the models were set up by first randomly sampling a stellar age \tstar and mass \mstar, and the parameters for the circumstellar material were then sampled randomly within ranges that depended on \tstar and \mstar. Thus, they broadly follow an evolutionary scenario. Recently \cite{Rob2017} published a new set of SED models where the parameters space has been uniformly sampled without assuming any relationship whatsoever between the various parameters, similarly to our present approach for the clump-level SED modelling. Indeed, very little is known from a quantitative viewpoint about the way a massive clump fragments and form stars, not only regarding the regimes of fragmentation and subsequent evolution of dense cores, but also regarding the feedback effects on clump dust temperature and the star formation history within each clump. We then believe it makes perfect sense at clump level to explore the widest possible parameters space and leave the various physical parameters completely independent from one another. When we turn to the modelling of a single YSO, however, we believe that our present knowledge about the formation of a single star is relatively speaking much more advanced to confidently assume that some of the physical parameters in a protostar+disk+envelope system may be related. In this sense modelling, say, a single YSO with a 1\msun\ protostar of age 1 Myr, plus a disk of 0.01\msun\ and an envelope of  20\msun\ might be interesting from an academic viewpoint but certainly would have little resemblance to a real physical system. We therefore believe it is perfectly legitimate to adopt the R06 SED YSO models grid rather than, e.g., the models of \cite{Rob2017} whose aim \textit{``is not to provide the most physically realistic models for young stars, but rather to provide deliberately simplified models for initial modelling''} (quoting the abstract). In other words, had we used the \cite{Rob2017} models we should have had to make some evolutionary assumptions to limit the areas of parameters space to astrophysically realistic models.

For the modelled protoclusters, for each compact source we searched through the R06's models set that which is the closest in mass and age using a Euclidean distance in $(\log{M_\star},\log{t_\star})$ space. Since the circumstellar properties of each model depend on \tstar and \mstar, compact sources that are more evolved will have on average an SED that also includes more flux at the shorter wavelengths. We stress that because the observable properties of the compact sources depend in this way on assumptions about the evolution of forming stars built-in to the R06 models (which include a dependency on stellar pre-main-sequence evolutionary tracks), the conversion to observable properties is only approximated. Since our intention is to model protoclusters, for which the fluxes will be determined over size scales larger than an individual young star, we use the fluxes in the largest aperture for the models (100,000\,au).

Rather than keep track of the SED over the full range of wavelengths, we instead use the SED defined by broadband fluxes calculated by convolving the SEDs over transmission curves (as described in \citealt{Robitaille+2007}) for the IRAC 3.6 to 8.0\microns, MIPS 24\microns, WISE 3.4 to 22\microns, MSX 8.3 to 21.3\um, PACS 70 to 160\microns,  SPIRE 250 to 500\microns, LABOCA 870\um, and BOLOCAM 1.1mm filters.

\begin{figure*}
\centering
\includegraphics[width=\textwidth]{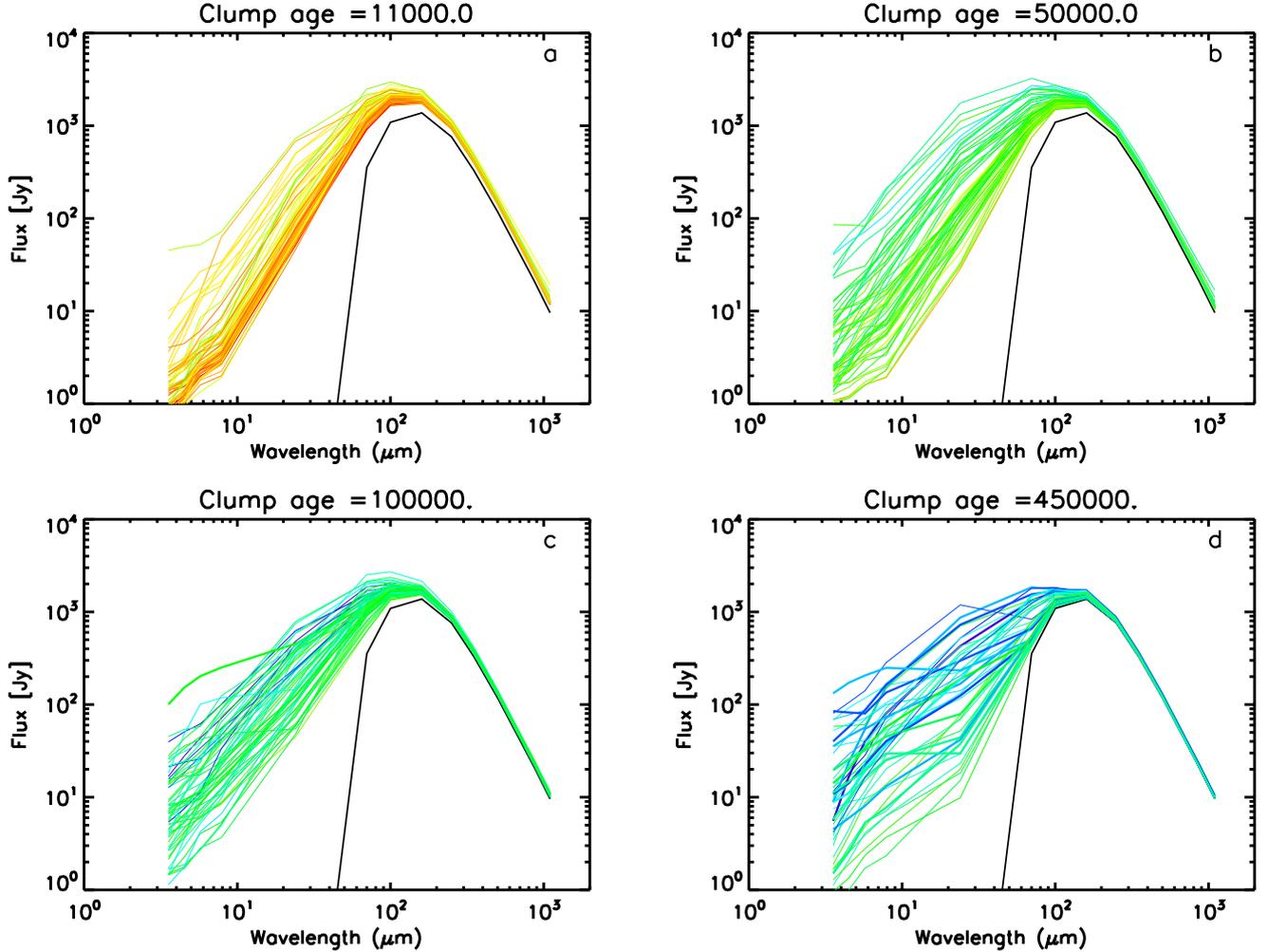}
\caption{\label{sed1} Sample SEDs from the model grid with \mclump =1000\msun, $ f_{core} = $0.3 and T$_{dust}$=20K. The four panels group models by age of the clump as indicated on the plots. The rainbow color coding from red to blue indicate the mean stellar mass in each synthetic cluster, spanning from 0.3 to 0.72\msun. Thick SEDs mark clusters where at least a B1 ZAMS star is present. The black line indicate the contribution of the clump diffuse dust. We note that since the values of the parameters are probabilistically extracted from a uniform distribution in the three parameters above, none of the models has \textit{exactly} the values specified above or we should have exactly 10 models for each panel of the figure. Models are instead selected within a few percent tolerance around those values. This also applies for other similar figures in the rest of the paper.}
\end{figure*}

With the SEDs of all compact sources in hand, the last component for the protoclusters model SEDs is the contribution from the diffuse inter-core cold dust of the clump. For this, we simply compute the modified blackbody with temperature \tdust assuming that the emission is optically thin, and using the dust properties for the $R_{\rm V}=5.5$ Milky-Way dust model from \cite{Draine2003}, which has a gas-to-dust ratio of 125, a C/H of 30 ppm and the grain size distribution "B" as defined by \cite{WD2001}.

An important effect of the clump dust not locked in star-forming cores is to extinguish the radiation emitted by the single YSOs systems sampled from the R06 models grid in the synthetic clusters. To statistically account for this effect, we have to make an assumption about the 3D distribution of clump dust and synthetic YSOs sampled. We model our clumps as radially symmetric envelopes with gas and dust volume density expressed as a power law of radial distance as:
\begin{equation}
\rho = \rho_0 [r/r_0] ^{-1.5}
\label{dens_law}
\end{equation}
valid from an internal clump radius r$_0$=0.01 pc \citep{Parmentier+2014} out to the outer clump radius. To avoid singularities the volume density is assumed constant at a value of $\rho_0$ for radii smaller than $r_0$. The exponent of the radial density power-law measured in dense clumps is between $-$1.5 and $-$2 \citep{Parmentier+2014}; although values between $-1.75$ and $-1.8$ are more frequently found using submillimeter continuum mapping \citep{Wu2010}, shallower profiles more similar to $-$1.5 are found in Herschel/Hi-GAL maps between 250 and 500\um\ where the extended emission at all scales is not filtered out as in ground based submillimeter observations. The external clump radius is parametrised as a function of mass as 
\begin{equation}
R_{Clump}=\left[ \frac{M_{Clump}}{2500M_{\odot}}\right] ^{1/2.4} pc
\label{rcl}
\end{equation}
obtained by fitting the mass-radius distribution of Hi-GAL protostellar clumps in \cite{Elia+2017}. Clump radii will then vary between 0.1 and $\sim$4.5pc over the \mclump	\ range explored in our models. For each synthetic YSO in the simulated embedded cluster we draw a random 3D position in the clump assuming a source radial density profile with a power-law exponent -2.5; this is steeper than the clump dust radial density profile (see Eq. \ref{dens_law}) following the prescriptions from \cite{Gutermuth2011} who analysed gas and YSO number density profiles in eight nearby star forming regions. We compute the line of sight extinction due to intervening clump dust toward the observer, adopting the same dust model from \cite{Draine2003} used to compute the clump dust emission, and then simply add the fluxes in different bands for the cold dust with the fluxes of all the compact sources to determine total fluxes in each band for each model protocluster. 

We remind again that in addition to our primary parameters mass and age, the R06 models include many more parameters like disk mass, outflow-driven envelope cavity amplitudes and density, and protostar line-of-sight orientation over which we randomly sample, implying that we do not impose any specific prescription or range. On the one hand, some of these parameters are implicitely limited to reasonable ranges in the R06 grid depending on evolutionary stage and total mass, reinforcing our choice to use the R06 models rather than the more recent 2017 models; on the other hand, parameters like orientation along the line of sight are left entirely free to simulate the physical random orientation of the sources in the cluster. To account for the variance of the selected individual R06 SED models for the parameters, as well as for the variance in the sampling of the mass function, we carry out 10 independent statistical realization for each primary parameters set,  giving a total of $2\times10^7$ synthetic protoclusters.

For each protocluster, and for the purpose of the analysis in subsequent sections, we determine the number of compact sources that contain stars that are consistent with being on or beyond the zero-age main-sequence (ZAMS). Since the compact sources rely on the R06 models, we use the stellar mass \mstar and age \tstar in each compact source along with the evolutionary tracks in R06 to determine whether a source has passed the ZAMS. 

\section{The properties of the model grid}
\label{analysis}

We now analyse the characteristics of the model grid in terms of SED shape and integrated luminosities throughout the entire parameters space explored. 

\subsection{Spectral Energy Distributions}


Figure \ref{sed1} provides a visual impression of the variance of SED shapes depending on some of the parameters. Models are selected with clump mass of 1000\msun, fraction of mass in cores of 0.3 and intra-cluster dust temperature of 20K, and are grouped in four different panels depending on the clump age as indicated on the figure panels. The different shapes of the SEDs at wavelength $\lambda \leq $100\um\ are due to the Monte Carlo sampling of stellar masses, ages and properties for the individual YSOs. We try to partially capture this effect by adopting a color coding based on the mean stellar mass in each cluster; the rainbow color coding goes from red (about 0.3\msun) to blue (about 0.72\msun). For this particular set of cluster models the total stellar mass goes from 130 to 300\msun\ while the number of synthetic stars spans the range between 300 and 650.

In each panel of Fig. \ref{sed1} the models have the same clump dust temperature, but the random sampling of YSO masses and ages generates, for each clump, clusters where variations of \mmean\ (the mean stellar mass) produce shallower SEDs with increased contribution of the mid-IR portion; even when not on the ZAMS, more massive YSOs produce (by Kelvin-Helmholtz  contraction and accretion) more power per unit mass and this is reflected in increasing amount of power radiated at $\lambda \leq 70$\um. The highest \mmean\ are reached for older clumps where stars had sufficient time to accrete from the clump material. Although the spread of SED shapes is apparently replicated for each of the four ages, the range of mid-IR fluxes spanned increases from  $\sim$3-6 Jy for the majority of the 10000yrs clump models, to $\sim$20-200 Jy for most of the 450000yrs clump models.


As we go to models with increasing clump age, the effects of the age sampling of the synthetic YSOs become more important as the YSO ages in a given clump are sampled with uniform probability from the models minimum age (10$^4$ yrs) to the assumed age of the clump. Older clumps can indeed be populated with relatively older and hence more luminous intermediate and high-mass YSOs, whose contribution is mostly apparent for $\lambda \leq$70\um. On average for clump ages above 10$^5$ years (Figs. \ref{sed1}c and d) the $\lambda \leq$ 100\um\ SED portion tends to be flatter than in younger clumps, again with a variance in flux levels that depends on the sampled YSO masses. For some of the models represented with thicker lines, at least one star is found to be on the ZAMS with a spectral class at least B1; for this particular model selection this happens for clump ages above 10$^5$ years.




\begin{figure}
\centering
\includegraphics[width=0.5\textwidth]{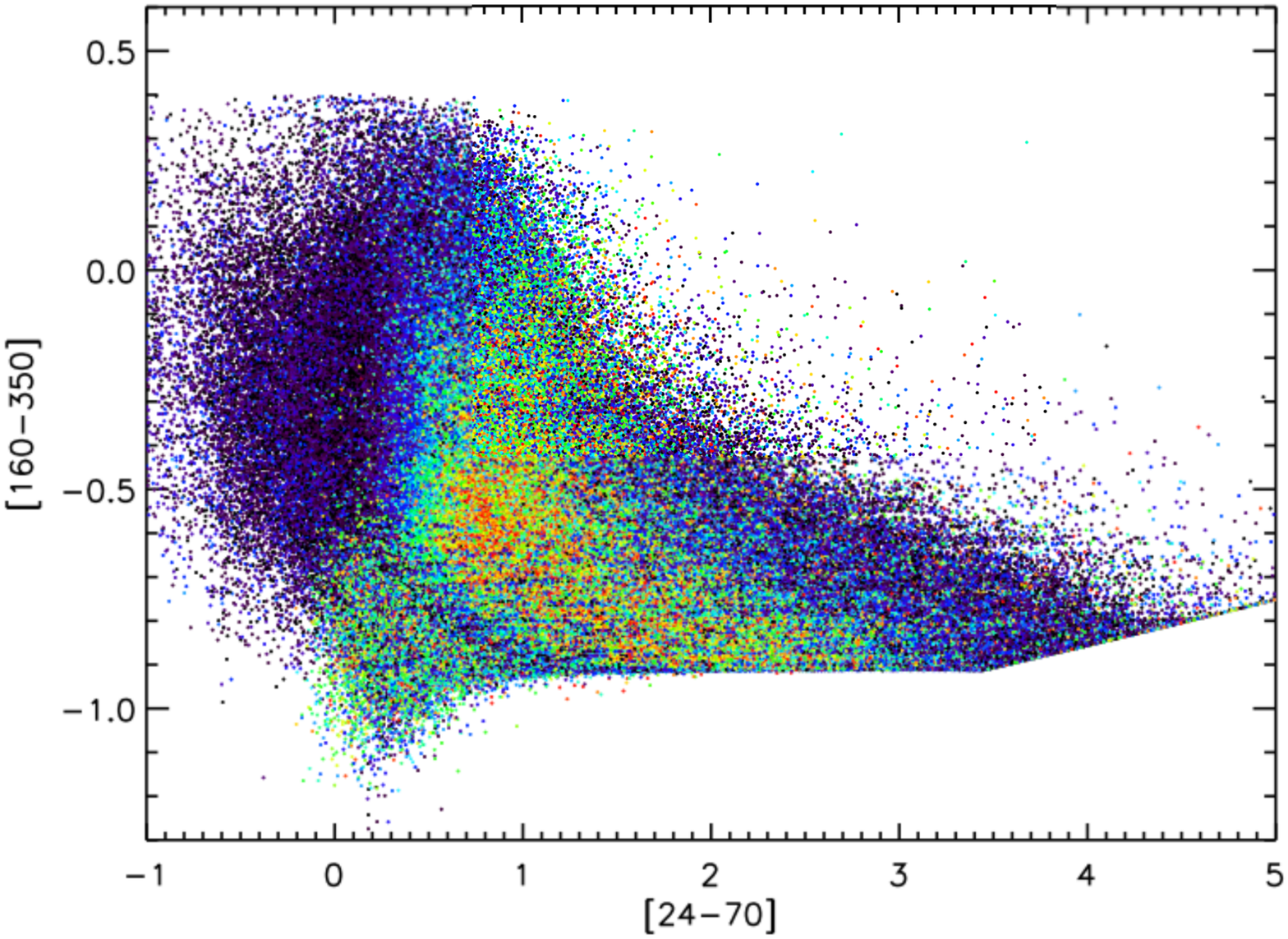} 
\caption{\label{colcol-age} Color-color diagram for all models, with points colour-coded by clump age from red (10$^4$ years) to blue-black (5$\times 10^5$ years).}
\end{figure}
\begin{figure}
\centering
\includegraphics[width=0.5\textwidth]{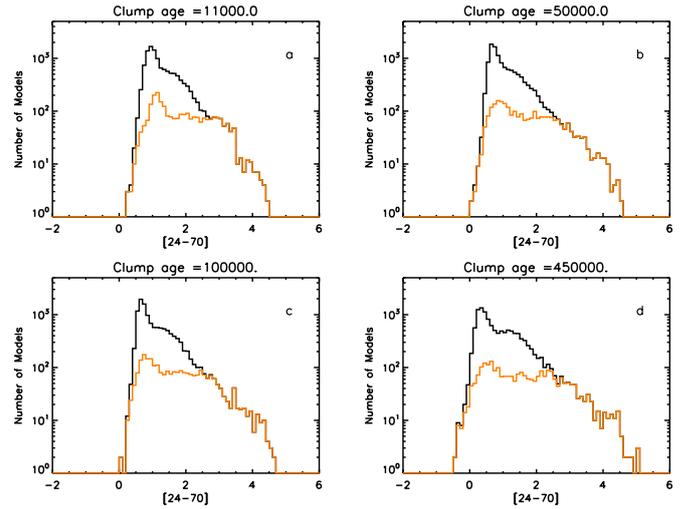} 
\caption{\label{col-age} [24-70] color histograms for models with \mclump =1000\msun\ and four clump ages. The color histograms represent the fraction of models with \fcore $\leq$0.1).}
\end{figure}

\begin{figure}
\centering
\includegraphics[width=0.5\textwidth]{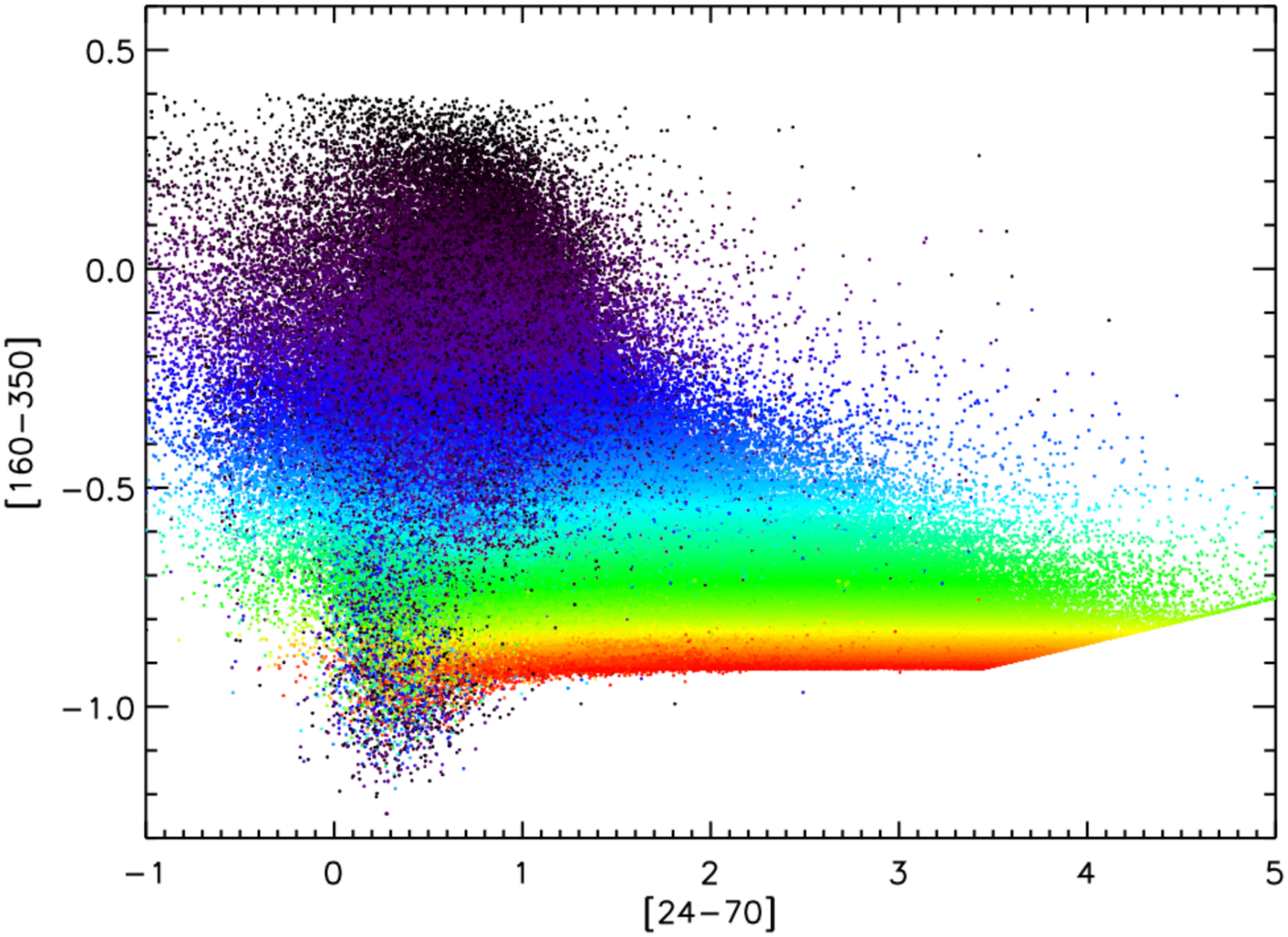}
\caption{\label{col-tdust} Color-color diagram for all models. Points are rainbow colour-coded by the clump dust temperature from red (10K) to blue-black (30K).}
\end{figure}

This is also illustrated synthetically for all models in Fig. \ref{colcol-age}, where this time the rainbow color coding is based on the clump age. Older clumps (blue-black points) are mostly concentrated toward low or negative values of the [24-70] color; as we proceed from [24-70]=0 to 1 we gradually find green and then red points. No obvious trend with age is instead observed for the [160-350] color, confirming that the latter is mostly influenced by the clump dust temperature. There is also a minority of the models extending up to [24-70]=4.5, with red to blue point intermixed and no clear indication of an age trend. Although, as we saw above in this subsection, a combination of relatively high clump dust temperature and deficiency of relatively high mass stars due to peculiar IMF sampling could  result in a steep 24-70\um\ SED irrespectively of the clump's age, the dominant cause for the swath of models with [24-70]$\geq 2$ in Fig. \ref{colcol-age} may reside in low fraction of mass locked in dense cores (and then available for star formation). 

Figure \ref{col-age} reports the [24-70] color distribution for all models with \mclump =10$^4$\msun\ for four clump ages as indicated; there is a decreasing trend in the peak of the [24-70] color distribution with clump age, but there is a scattered distribution of models at [24-70]$\geq 2$ for all ages that is dominated by models with \fcore $\leq 0.1$ (the orange histogram). It is plausible that when the large majority of clump material is not locked in dense cores and then available for star formation, the total emission at $\lambda \geq$70\um\ is dominated by the clump dust and not by YSOs emission, making the [24-70] color relatively insensitive from the properties of the YSOs population (including age).

Clump mass and dust temperature are the parameters that almost uniquely determine the SED for $\lambda \geq$100\um. This can be seen in Fig. \ref{col-tdust} where the FIR colours for the entire grid of models is represented. Each point represent a model and the points are colour-coded by their clump dust temperature; this parameter is the one that almost uniquely determine the appearance of the clump SED longward of 100\um. The lack of models in the top-right area of the plot is due to the incompatibility of steep [24-70] colours with rising \tdust.

\begin{figure}
\centering
\includegraphics[width=0.5\textwidth]{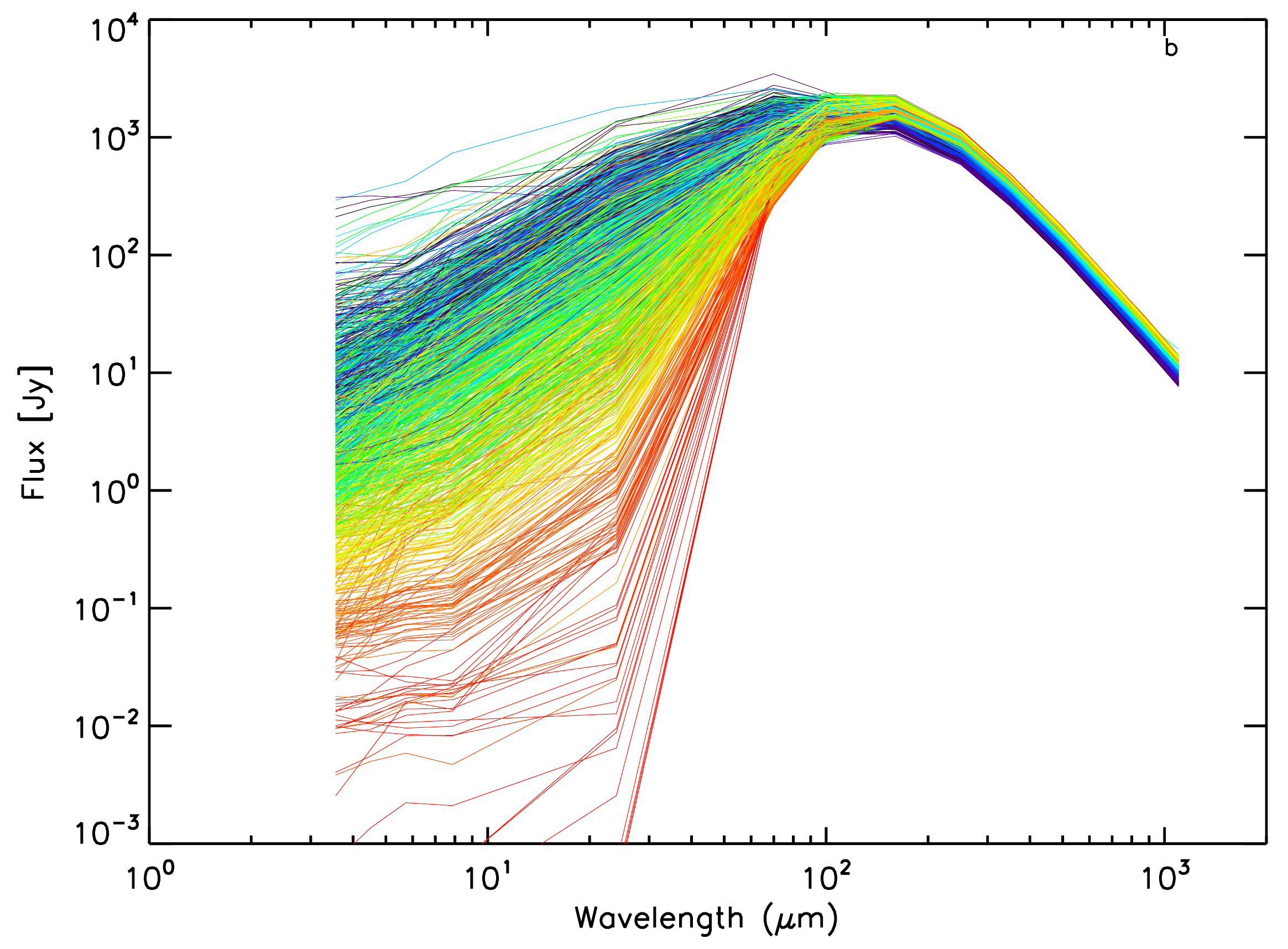}
\caption{\label{sed3} Sample SEDs from the model grid with \mclump =1000\msun, clump age of 10$^5$ years, and dust temperature of 20K. SEDs are colour-coded by $f_{core}$ (clump mass fraction in dense cores) ranging from red (0.1) to blue (0.5); thick lines indicate models where at least on YSO is a ZAMS star.}
\end{figure}

Finally, we show the effect of increasing the fraction of clump mass locked in cores. Figure \ref{sed3} reports a selection of models with clump mass again =1000\msun\, but with clump age fixed at 10$^5$ years and clump dust temperature of 20K. The color coding of the SED is based on $f_{core}$, and shows that increasing the fraction of mass locked in cores has the  effect of flattening the portion of the SED below 100\um\ in a way that seems qualitatively similar to the effect of clump age. 

\begin{figure}
\centering
\includegraphics[width=0.5\textwidth]{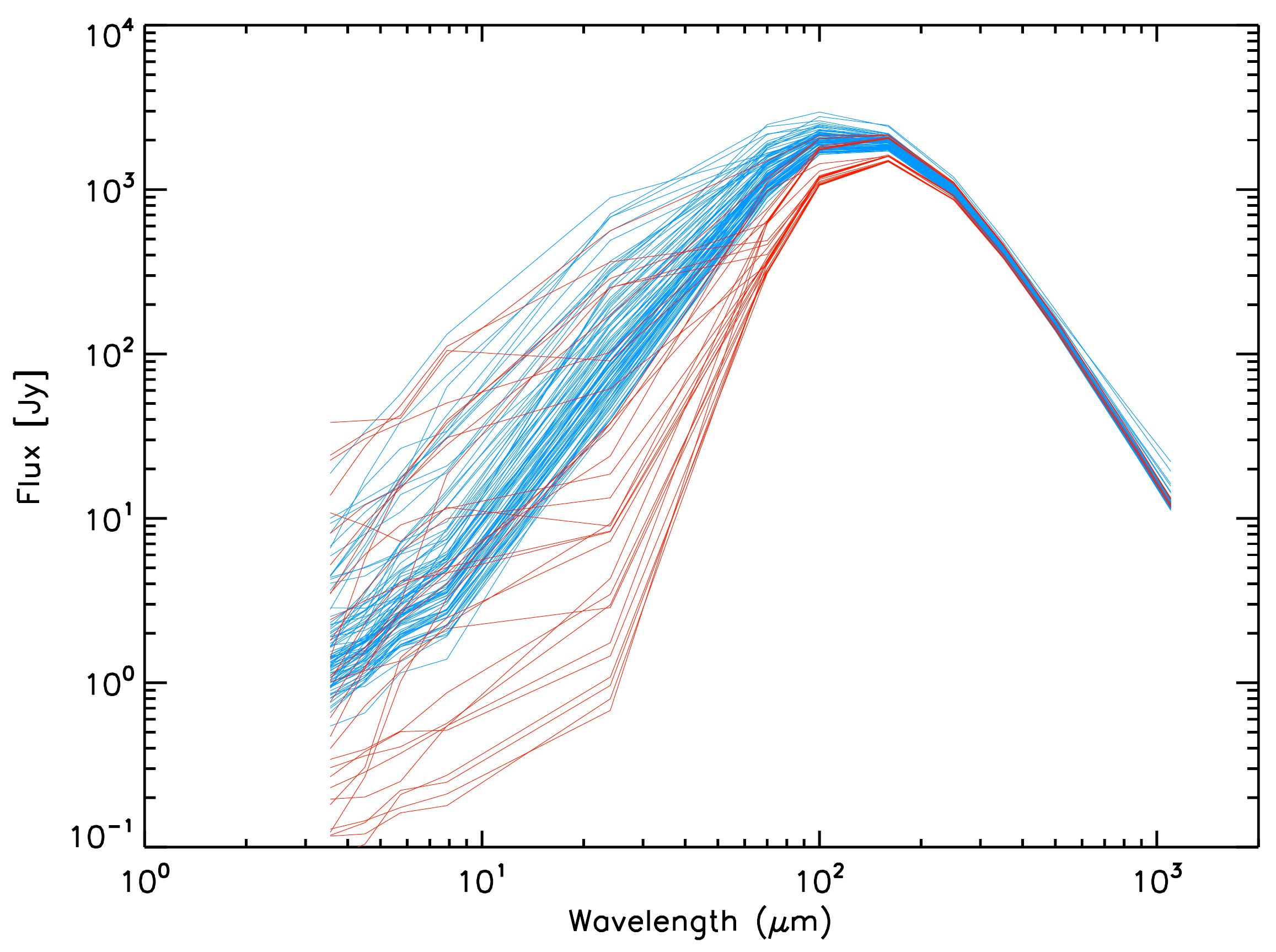}
\caption{\label{sed_degeneracy} Sample SEDs from the model grid with \mclump =1000\msun, and \tdust\ 20K. Blue SEDs are for clump age of 10$^4$ yrs and \fcore =0.3, while red SEDs are for clump age of 5$\times 10^5$ years and \fcore = 0.1.}
\end{figure}

To analyze whether an age-\fcore\ degeneracy is affecting the model grid, we show in Fig. \ref{sed_degeneracy} two groups of SEDs for clumps of mass 1000\msun and T$_{dust}$=20K that differ for the combination of clump age and \fcore. To see the combined effects of clump age and \fcore\ at work in this model set, we report in blue the subset of models for very young clumps and intermediate \fcore\  and in red the subset with very old clumps but low \fcore. Low \fcore\ models (red) for a given clump mass cannot match the amount of power radiated by intermediate \fcore\ (blue) even for relatively old ages. In the first case the SEDs are steeper than in the latter case between 24 and 100\um, while the SED portion is much flatter below 24\um. We are led to believe that the detailed analysis of the SED shape at $\lambda \leq$100\um\ allows to disentangle the effects of clump age and \fcore.

\subsection{Integrated properties and evolutionary diagnostics}

\begin{figure*}
\centering
\includegraphics[width=0.49\textwidth]{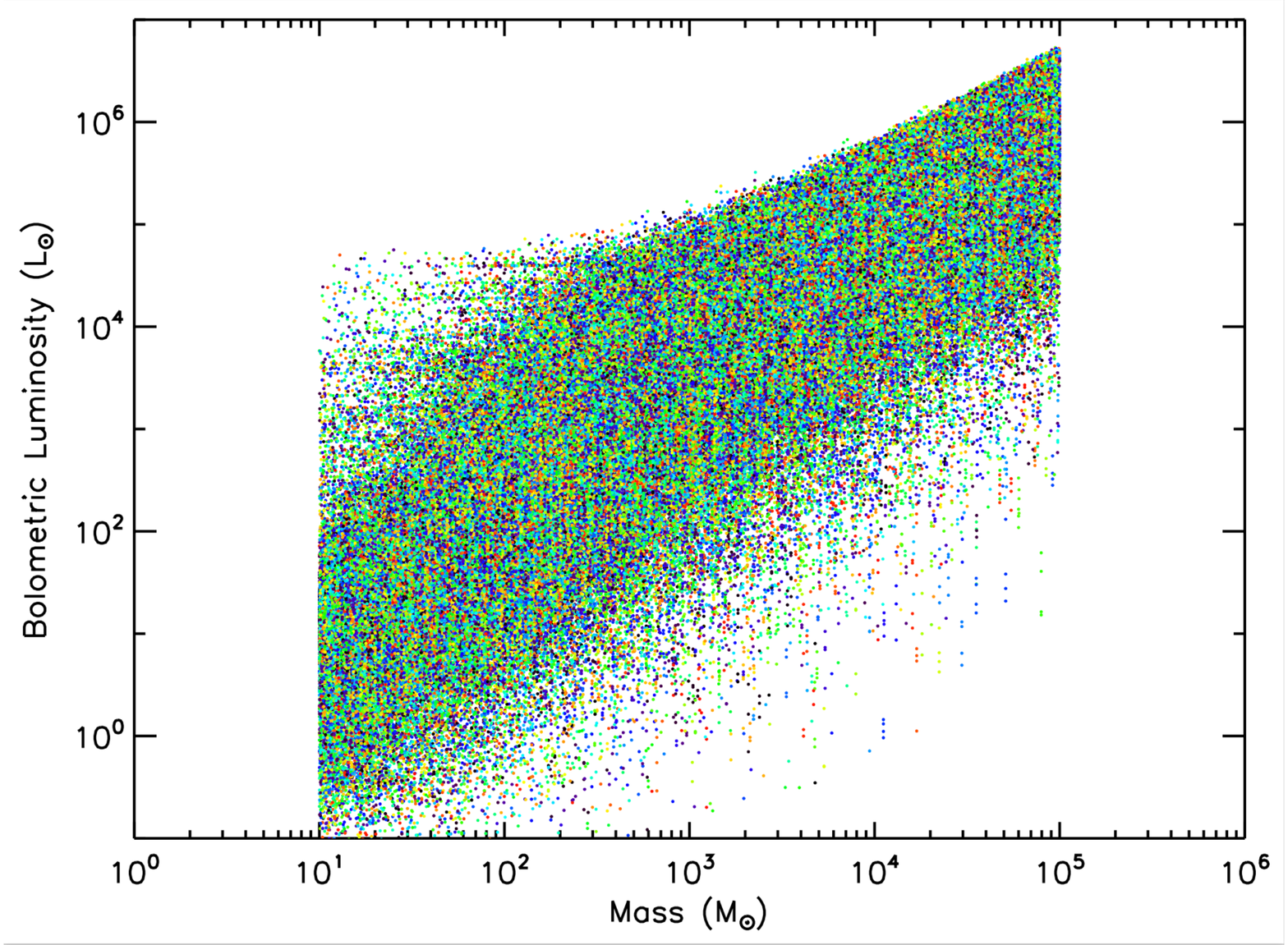}
\includegraphics[width=0.49\textwidth]{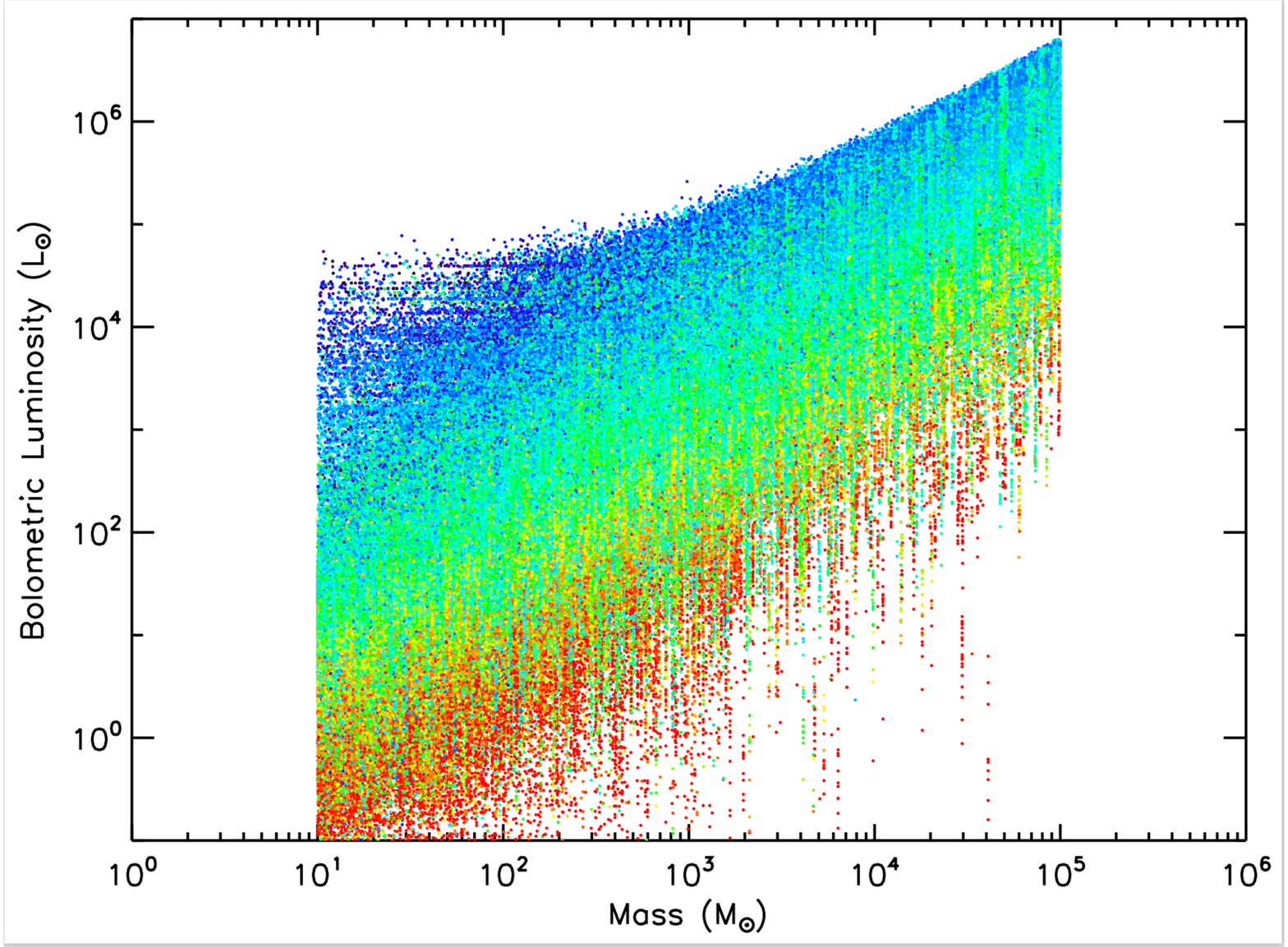}
\caption{\label{lm1} \loverm\ plot for all models in the grid. Rainbow colour-coding is by clump dust temperature for the left panel (10$\rightarrow$30K from red to blue), and by clump [24-70] color in the right panel (4$\rightarrow$-1.5 from red to blue).}
\end{figure*}

We turn to analyse the properties of the cluster model grid as a function of integrated quantities like mass and luminosity. As we have seen in the previous sections, unless \fcore\ is below 0.1 the most energetic portion of the models SED ($\lambda \leq$100\um) is dominated by the properties of the YSO population in the synthetic clusters. To verify how this reflects in the luminosity, we present in Figure \ref{lm1} the L/M diagram for all models, where different model colour-coding are adopted for clump dust temperature and clump [24-70] color in the two panels. The luminosity is obtained integrating the models SED from the IR to the millimeter, so it is not properly speaking the bolometric luminosity; the IR-mm wavelength range dominates the SED for a large fraction of the models, but as Fig. \ref{sed1} shows, there is a significant fraction of models where the $\lambda\leq$3\um\ portion (that we do not cover here) can make an important contribution. Unless otherwise noted in the rest of the paper, by luminosity L we then intend the integral over the model SED, and by M the total clump mass. The left panel shows no clear segregation of model colours, suggesting that no dependence of clump luminosity for any given clump mass is apparent with the clump dust temperature. We have to remember here that our approach is to uniformly sample the entire parameters space with no \textit{a priori} physical plausibility filtering; as such, there are for example cases of relatively high L/M (suggesting for the majority of models the presence of a protocluster of luminous YSOs) coupled with very low \tdust\ that would not be produced had we performed radiative transfer computation to link the radiative output of the YSOs to the clump dust.

On the other hand the right panel of the figure clearly shows that for any given clump mass the distribution in bolometric luminosity is determined by the [24-70] color. The transition from red to blue points (decreasing [24-70] values) is not sharp and reflects the influence by other model parameters like \fcore\ and peculiarities in the IMF and age sampling of the individual YSOs in the synthetic clusters, but the overall trend is clear. This result, together with the indication of the [24-70] color dependence on clump age, shows that both luminosity and [24-70] colour are good probes for the evolutionary stage of dense massive clumps undergoing star formation confirming via more statistically robust modelling our earlier proposal \citep{Molinari+2008} based on analytical models for the formation of a single high-mass YSOs in dense clumps.

\begin{figure}
\centering
\includegraphics[width=0.5\textwidth]{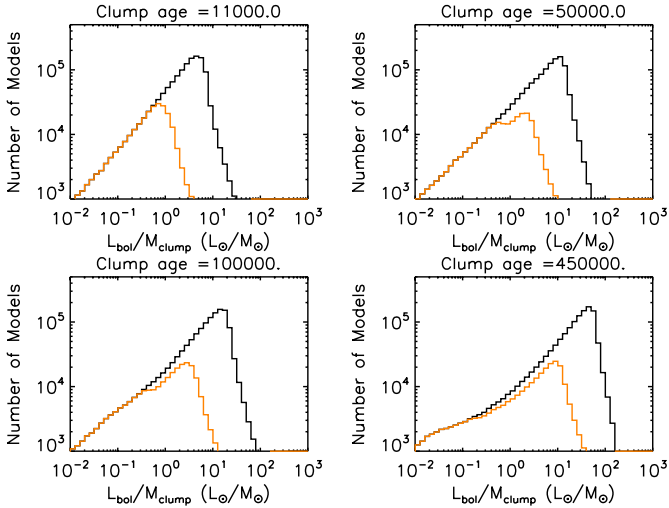}
\caption{\label{lm-age} Distribution of \loverm\ models for four clump age as indicated on the panels. The orange histogram represent the distribution for all models with \fcore $\leq$0.1 }
\end{figure}

As a convenient evolutionary indicator we propose the \loverm\ ratio between the clump bolometric luminosity and its mass (that we will abbreviate in L/M, in units of \lsun /\msun). The distribution of L/M as a function of clump age is presented in Fig. \ref{lm-age}. The peak of the distribution of models clearly and significantly peaks toward higher L/M as the clump age increases. The extended tail of the distribution at low L/M is dominated by models with \fcore $\leq$0.1 (orange histograms). Any attempt to generate evolutionary tracks from this extensive grid of models will have to take this parameter into account (see \S\ref{tracks}).

\begin{figure}
\centering
\includegraphics[width=0.5\textwidth]{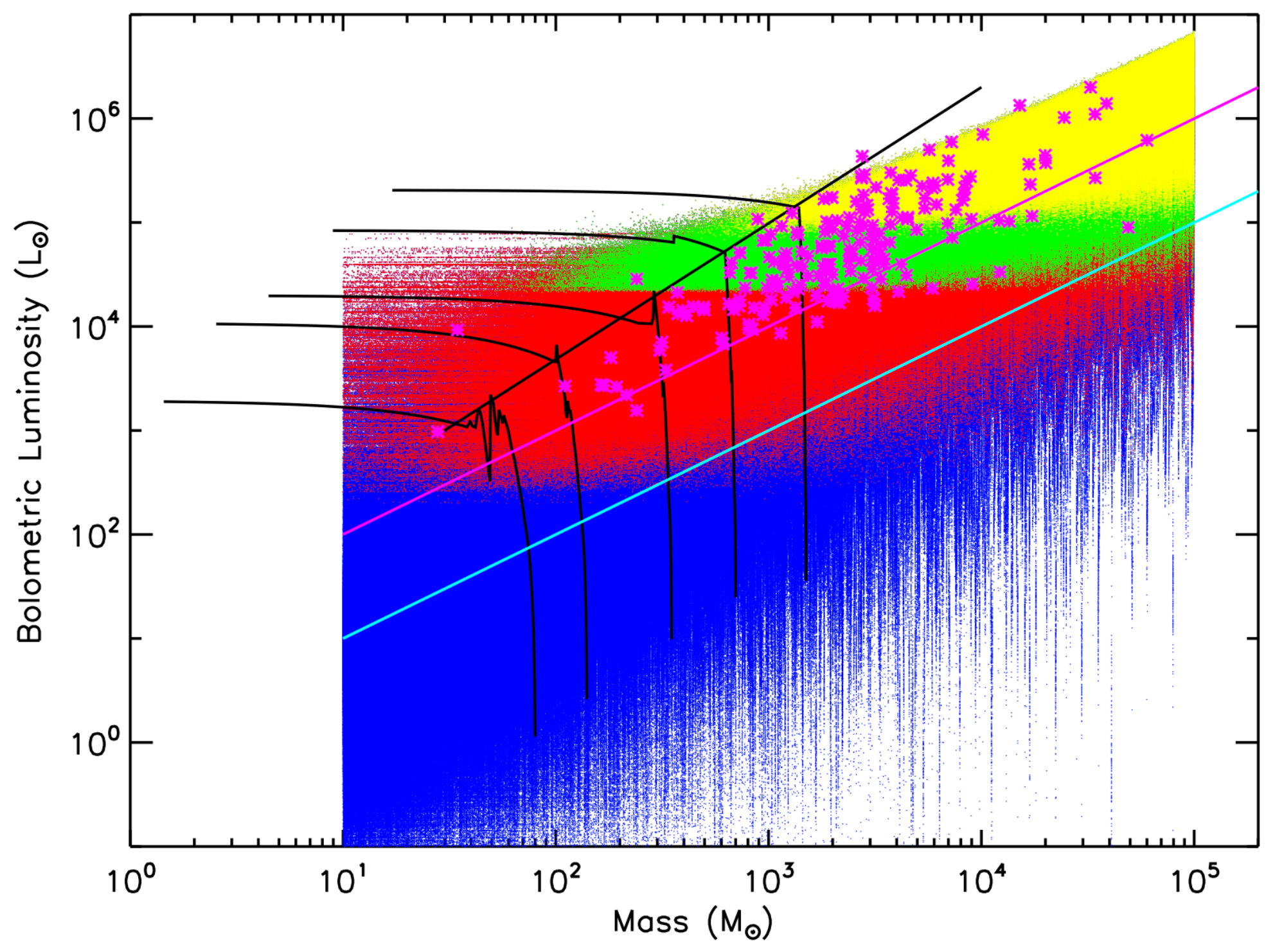}
\caption{\label{lm2}  L/M plot for all models in the grid (blue points). Models in which at least one YSO is on the ZAMS are indicated with a red dot, and those in which the ZAMS YSO is of spectral class B0.5 or earlier is represented by green dots. Models in which there are at least 3 ZAMS stars earlier than B0.5 are indicated by the yellow dots. The cyan and magenta straight lines are from \protect\cite{Molinari+2016c}, and the magenta stars  represent the position occupied by the dense clumps in the \higal\ clump catalogue \protect\citep{Elia+2017} with a \hii\ region counterpart from \protect\cite{Cesaroni+2015}. The black lines represent the evolutionary tracks from \protect\cite{Molinari+2008}, together with the location of \hii\ regions from the same work.}
\end{figure}

We now compare in Figure \ref{lm2} the location of the model clusters with respect to the evolutionary tracks that we computed in \cite{Molinari+2008} assuming that a single intermediate/high-mass star was forming in the dense clumps. The ascending portion of the tracks (black lines) marked the main phase of the accretion of the massive YSO that was rapidly gaining luminosity while accreting mass at an ever increasing accretion rate, according to the model of \cite{mck03}; the tracks were stopped at the location occupied by UCH{\sc ii} regions (the oblique black line at the end of the ascending portion of the tracks), whose luminosities were computed based on IRAS fluxes, and that we dubbed as the ``ZAMS line''. 

The models computed in the present work are much more plausible as they assume that a cluster of YSOs is forming in the dense clump, rather than a single star. Contrary to the analytical models in \cite{Molinari+2008}, we are not here following the evolution of each clump, as each clump results from an independent Monte Carlo realization for each cluster parameter set; rather, the evolution can be followed in a broader sense by mapping the age of the clusters as all other parameters vary. The L/M plot is again convenient as we proved that in the present model grid the bolometric luminosity is a good average tracer of the clump age. All models are represented by the blue points in the figure; among these, the models where at least one of the YSOs is a ZAMS star are indicated with red dots; as they on average represent more evolved clusters, it is not surprising that they are concentrated towards the upper portion of the models distribution. The lower boundary of the distribution of these model clumps (the red dots) agrees very well with the L/M=1 line (see below) down to $\sim$200 \msun, and then flattens at lower masses; this is due to the upper limit that we set for the clump age \sftime$\leq 5\,10^5$ years, that is too short for low mass (and hence lower luminosity YSOs) to reach the ZAMS. The subset of models where this ZAMS object is of spectral class B0.5 or earlier is indicated with green dots; such models occupy the topmost portion of the distribution in the plot. Finally, we report in yellow dots the models where at least three ZAMS stars are found. The density of dots in the plot is such to fill completely the white space and each color covers the corresponding area; we will return later on this point.

The resulting scenario confirms that the bolometric luminosity of dense clumps is an excellent tracer of the evolutionary stage of the ongoing star formation, with noticeable improvements with respect to the very simplified picture of \cite{Molinari+2008}. Models show that when clusters are considered instead of single high-mass stars, individual members may reach the ZAMS earlier during the period when the clump increases its luminosity; the exact location where this happens in the L/M plot depends on the specific parameters of the models, but the increasing frequency of the red dot symbols with luminosity suggests that as clump age proceeds, this will happen independently of the other parameters. It is interesting that in a recent survey \citep{Molinari+2016c} of star-forming clumps in \cht, which is a temperature probe for dense gas in star-forming cores, clumps were being detected with a nearly 50\% rate as soon as  L/M$>$1, with a minimum gas temperature of $\sim$35K. This value is represented with the cyan line in Fig. \ref{lm2} and nicely corresponds to the lower envelope in the distribution of the red dot symbols. 

As evolution proceeds the clumps gain more luminosity and will have a higher chance of forming more massive ZAMS star; the lower envelope of the distribution of the clusters hosting at least a B0.5 ZAMS star (the green dot symbols) is situated at L$\sim 2\,10^4$\lsun. The magenta line that corresponds to L/M=10 marks the threshold where in the above mentioned \cht\ survey the detection rate reached 100\%, and where the gas temperature was found to rise as a power-law of the bolometric luminosity as expected in case of a strong embedded energy source like an intermediate or high-mass ZAMS star. It is also remarkable how dense clumps in the \higal\ survey that were found associated with a H{\sc ii} free-free CORNISH radio source \citep{Cesaroni+2015}, the magenta asterisks  in Fig. \ref{lm2}, are consistently found in the same region covered by the cluster models hosting a B0.5 ZAMS star or earlier. Some of these H{\sc ii} regions for M$\leq 2000$\msun\ are also found at luminosities below the threshold for a B0.5, indicating that a later class ZAMS star is present.

\begin{figure}
\centering
\includegraphics[width=0.5\textwidth]{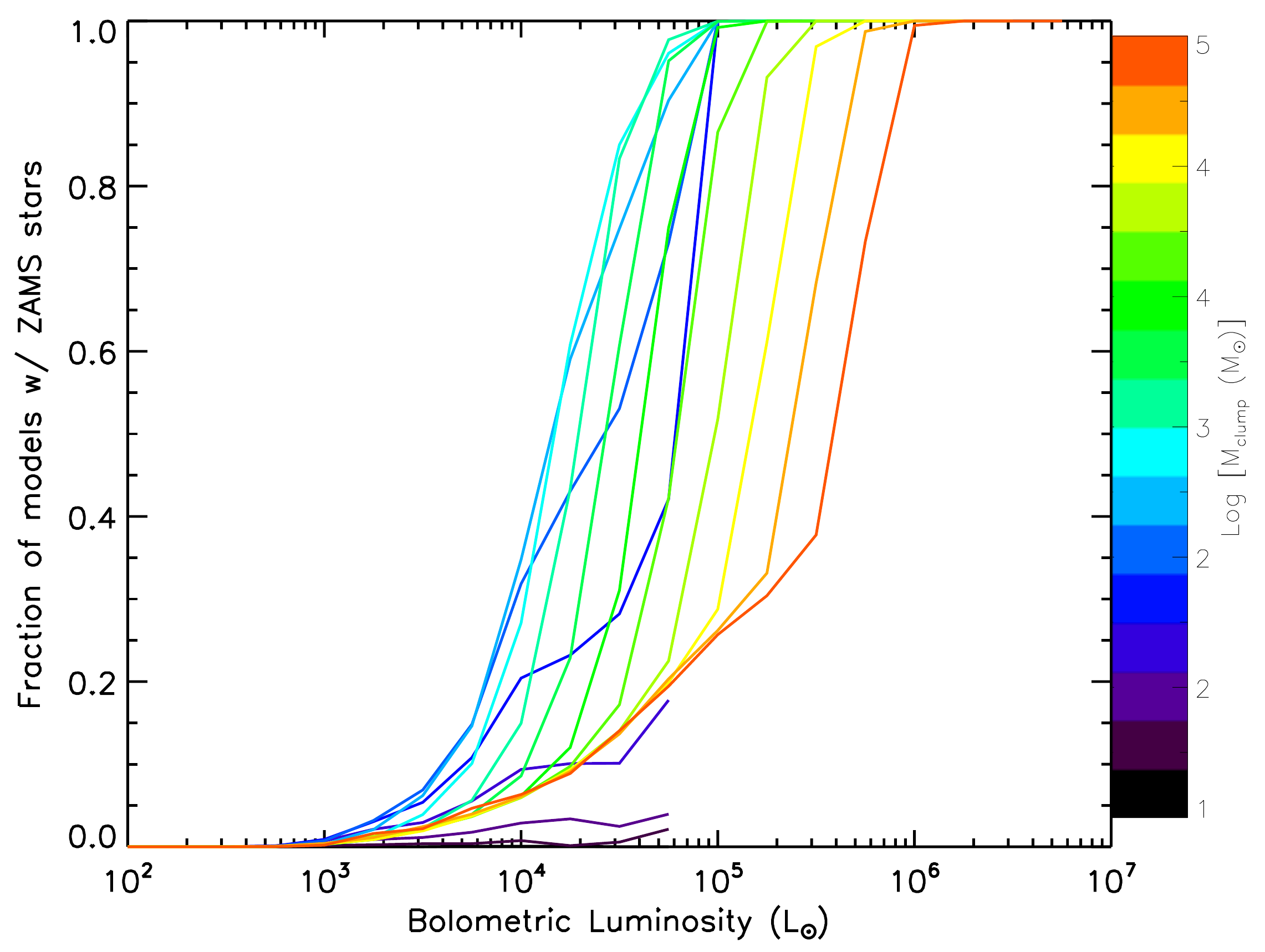}
\caption{\label{zamsfrac}  Fraction of models containing at least one ZAMS star over the total, as a function of bolometric luminosity for clump of different masses represented by different colours. The clump mass varies from 10 to 10$^5$\msun\ in logarithmic intervals of 0.25.}
\end{figure}

The concept of "ZAMS line" in the L/M diagram should therefore be substituted with the concept of a "ZAMS strip". Synthetic cluster models as well as observations concur to define a quite plausible evolutionary scenario where massive clumps can effectively be followed along their evolution thanks to global parameters like the bolometric luminosity (or the \loverm\ ratio) and the clump-integrated [24-70] color.

Fig. \ref{lm2} shows that there are clump mass-dependent luminosity thresholds where clumps hosting a ZAMS star, a B0.5 ZAMS or three ZAMS stars are found; the presence of clumps in those areas of the L/M plot is a necessary condition for them to host such stars. It is however, not a sufficient condition. The density of symbols in the plot is such that for example it cannot be ascertained whether blue-dot cluster models are present in the area covered by the other symbols. We then computed the 2D models density distribution in the L/M plot in square  logarithmic bins of 0.25; Fig. \ref{zamsfrac} reports for each clump mass bin (represented with a different color) the fraction of models hosting a ZAMS star (i.e., the fraction of blu-dots under the red area) as a function of the bolometric luminosity. The figure shows that even for a clump of 1000\msun\ it is possible to find models with no ZAMS even for luminosities up to 10$^5$\lsun. 

\begin{figure}
\centering
\includegraphics[width=0.5\textwidth]{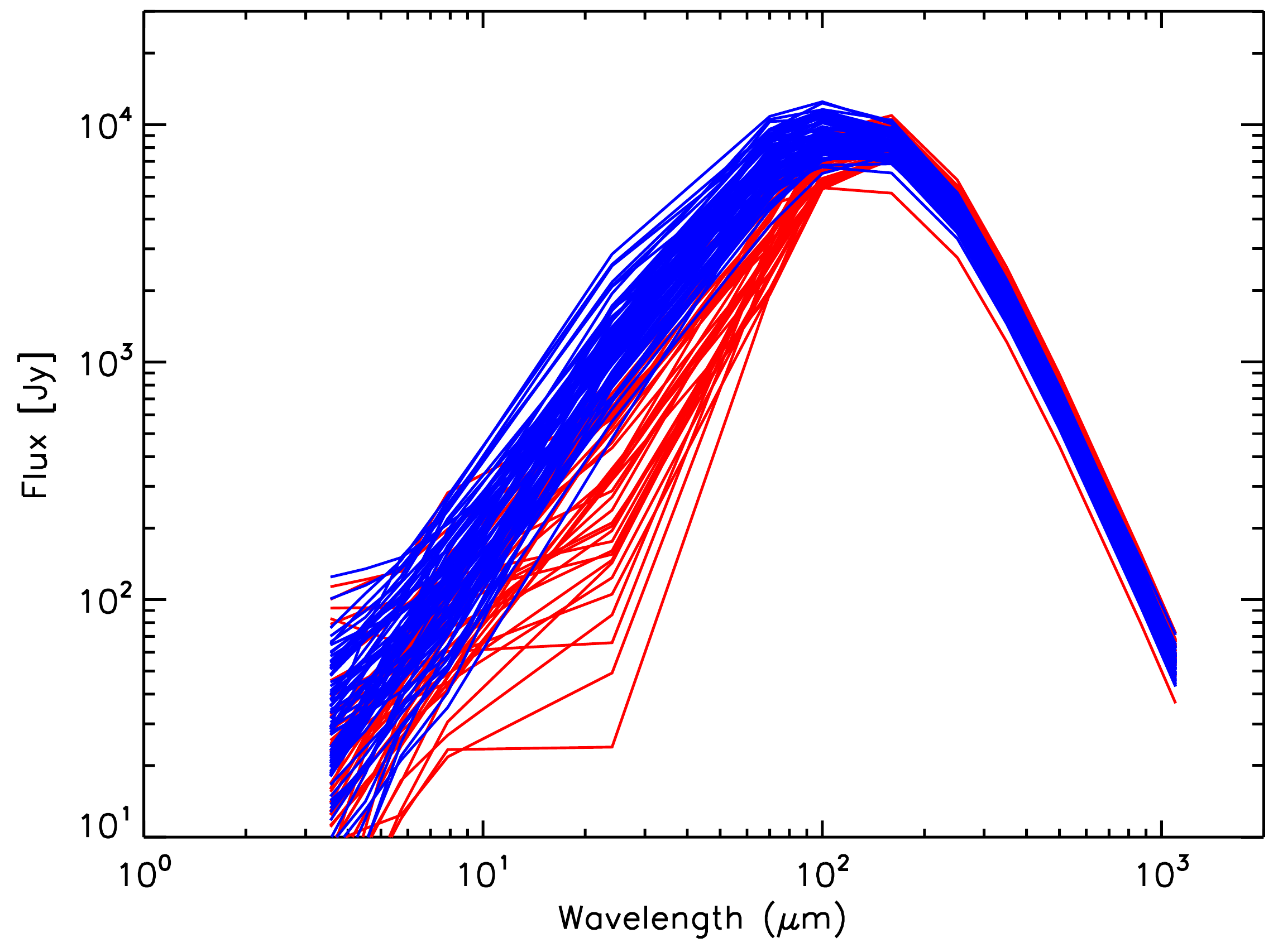}
\includegraphics[width=0.5\textwidth]{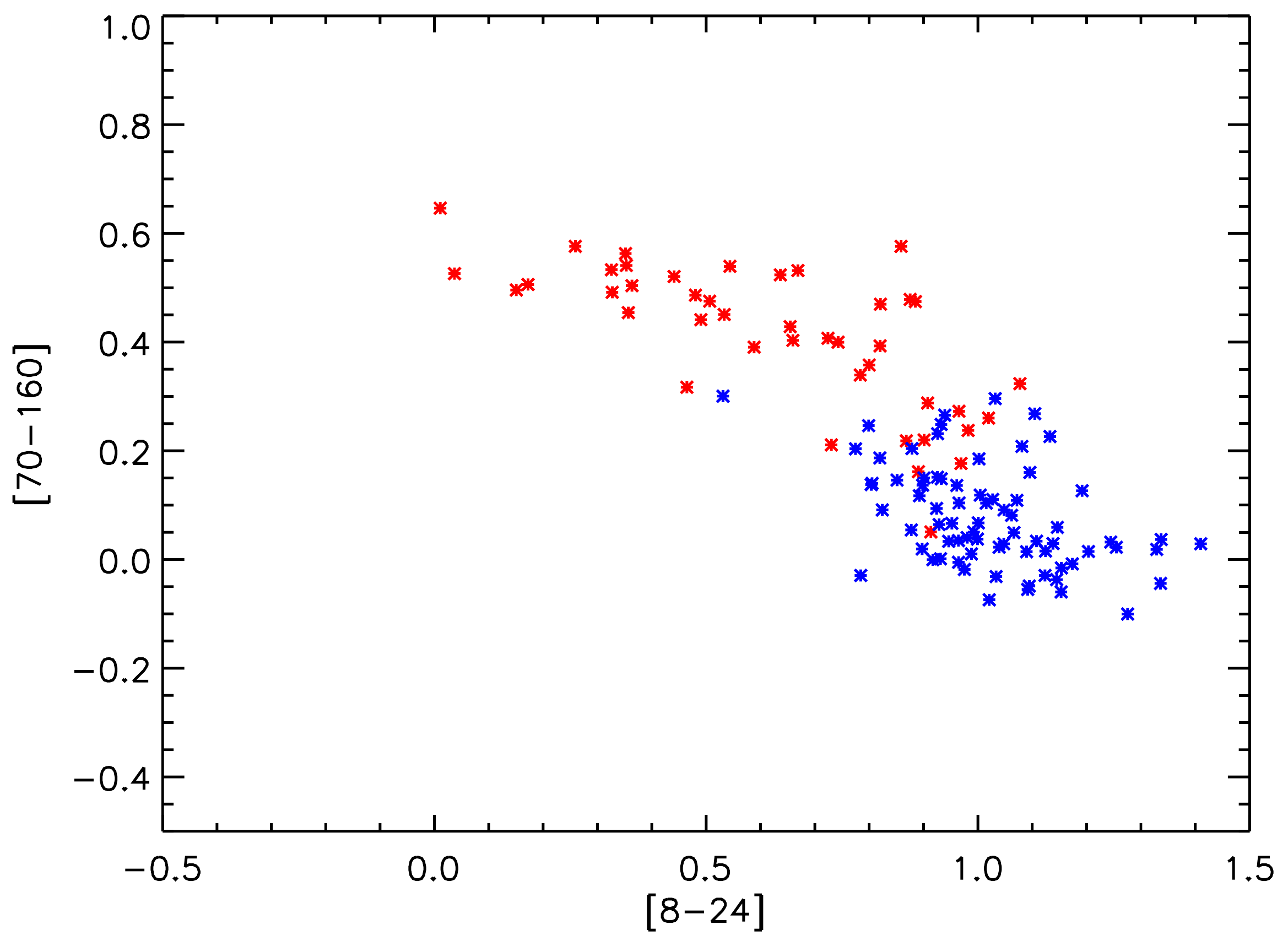}
\caption{\label{zamsnozams} \textit{Top: } SEDs for model sets with \mclump =5000\msun, T$_D$=20K, L$_{bol}$=50000\lsun\ and with at least one ZAMS star in the model clusters (red SEDs), and without ZAMS stars in the clusters (blue SEDs). \textit{Bottom: } [70-160]\textit{vs}[8-24] diagram for the SEDs of the left panel, with the same color coding.}
\end{figure}

It is very interesting then to understand what are the properties of such model cluster where no ZAMS are found in spite of their very high luminosity. Fig. \ref{zamsnozams} (top) reports the SEDs for model clusters with clump mass of 5000\msun, dust temperature of 20K and luminosity of 5$\times 10^4$\lsun; color coding is based on the presence (red) or absence (blue) of at least one ZAMS star. The shapes of the SEDs of clusters without a ZAMS star (blue) are shallower, peaking at $\sim$100\um\ and then slowly decreasing toward shorter wavelengths. SEDs of clusters with at least a ZAMS star (red) instead peak at slightly longer wavelengths, decreasing more steeper than the blue SEDs and then flattening more as they reach the 20\um\ region. The wavelength ranges where the SEDs steepness is different are mostly around the peak and below 20\um; this is well represented in the [8-24] $vs$ [70-160] color plot in the bottom panel of Fig. \ref{zamsnozams}. The two classes of models substantially differ by fraction of clump mass locked in cores and clump age, with median values of [\fcore, Log(age)] = [0.14$\pm$0.1, 5.25$\pm$0.33] for the models with ZAMS stars, and [0.43$\pm$0.07, 4.54$\pm$0.23] for models without ZAMS stars. The very different fraction of clump gas locked in cores is reflected in a stellar content (both in numbers and in total mass) that is three times higher in the latter models compared to the former and explains the similar energy output budget. The stars in the latter class of models are younger. A similar situation is found for other clump mass and luminosity ranges and clump dust temperature.

Different clump masses, luminosities and clump dust temperatures \tdust\ will select model SEDs for clusters with and without ZAMS stars that will differ from those used in Fig. \ref{zamsnozams}. We verified that the [70-160] color is degenerate on these three parameters but the [8-24] is not, and independently from the mass of the clump, its luminosity and dust temperature, the distributions of [8-24] color between clusters with and without ZAMS stars are always distinct as in Fig. \ref{zamsnozams}(bottom), with a transition region 0.8$\lsim$[8-24]$\lsim$1.1.

\subsection{Physical plausibility of the synthetic models}
\label{balance}

As anticipated in \S\ref{modelsgrid}, our synthetic models do not make any \textit{a priori} assumption about the relationship of temperature of the intra-clump dust \tdust\ with the energy input provided by the embedded simulated YSOs. As the radiation emitted by the embedded YSOs is a primary source of energy to heat the clump medium, we would expect that any model with a significant unbalance between the fraction of YSOS-radiated luminosity absorbed by the intra-clump dust and the continuum luminosity emitted by it should be deemed as unphysical. We will first investigate the impact of such occurrences in our grid of models, and then provide simple quantitative arguments for a number of reasons why this may not be the case.

The total input power emitted by the YSOs population that is effective to heat the intra-clump dust is less than the total YSOs bolometric luminosity from the Robitaille's models, as the fraction of radiation emitted by YSOs at wavelengths where the intra-clump dust is significantly thin will not be reprocessed. To obtain a crude estimate of this threshold wavelength as a function of the clump mass we recall that
\begin{equation}
\tau (\lambda)=C_{ext}(\lambda) N_H
\label{tau}
\end{equation}
where we take the extinction coefficients for the same dust mixture we use in the models (see \S\ref{modelsgrid} above) and where we have neglected scattering, as we are exploring ranges beyond 20\um. The total column density can be obtained by integrating the volume density along a radial direction from the inner clump radius (r$_0$=0.01 pc) up to the outer radius R$_{Clump}$ (see Eq. \ref{rcl}), assuming the gradient of Eq. \ref{dens_law}. The total column, in the limit R$_{Clump}>>$r$_0$ that is always verified in our models grid, can be expressed as 
\begin{equation}
N_H\sim 2\rho_0 r_0
\label{nhtot}
\end{equation}
The value of $\rho_0$ as a function of clump mass is again obtained integrating the clump mass radial profile and is
\begin{equation}
\rho_0={{1.5 M_{Clump}}\over{4 \pi r_0^{1.5} \mu _H R_{Clump}^{1.5}}}
\label{rho0}
\end{equation}
where R$_{Clump}$ is parametrised as in Eq. \ref{rcl} and $\mu_H$, adopted as 1.4 \citep{Kauffmann+2008}, is the mean gas mass per unit H. Using (\ref{rho0}) in (\ref{nhtot}) and substituting in (\ref{tau}), we can solve for $\lambda (\tau=1)$. As an example for M$_{Clump}$=[10$^2$, 10$^4$, 10$^5$] \msun\ one obtains $\lambda (\tau=1) \sim$ [30, 100, 140]\um.

For all models in the grid we are then in the condition to compute the fraction of radiation emitted by the YSOs that is below $\lambda (\tau=1)$, $L_{\star MIR}$, that can be absorbed by intra-clump dust resulting in its heating. The estimate of the amount of input internal luminosity that is reprocessed in the clump, however, critically depends not only on the average column density of intra-clump dust, but also on its degree of "clumpiness" (i.e. its filling factor), on the specific dust opacity adopted and the spatial distribution of irradiating sources within the clump. The variance of conditions that can be found is such that a factor 2 possible variation in the total luminosity radiated by the embedded population of YSOs is not important. A proper account of these effects would require, again, a full spatially resolved radiative transfer treatment which is beyond the scope of this paper and cannot certainly be carried out for a parameters space as large as the one presented in this paper. What we want to derive here is a "rule-of-thumb" metric to flag and parametrize conditions where models could be inherently energetically unbalanced, and offer possible explanations of why such unbalances could be due to specific conditions not accounted for in our simple modeling.

The luminosity emitted by the intra-clump dust, L$_{C, T_D}$, can be easily computed for each model by integrating a modified blackbody at temperature T$_D$ emitted by a mass \mclump (1-\fcore) with the same dust opacity (\citealt{Draine2003}) used for YSOs SED models. However, one should consider that molecular clumps are also externally heated by the ambient Interstellar Radiation Field (ISRF); its intensity greatly varies both globally, for different locations in the Galactic Plane, and locally due to the proximity of ionising sources like \hii\ regions or OB associations. To qualitatively estimate this effect we look at the average properties of the so-called ``pre-stellar'' clumps in the Hi-GAL survey. These are compact clumps with no flux detected below 100\um\  that can be very-well modeled with single-temperature greybody emission with no clear evidence of an internal energy source, differently from the ``protostellar'' clumps that instead show clear emission below 70\um\ down to the mid- and near-IR. \cite{Elia+2017} reports an average temperature of 12K for pre-stellar clumps and we will here assume that on average this is the temperature reached by clumps due to irradiation by the ambient ISRF \citep{Wilcock+2011}, with a corresponding radiated far-IR power L$_{C, 12K}$. When we compare $L_{\star MIR}$ with [L$_{C, T_D}$-L$_{C, 12K}$], we find that they are in agreement within a 50\% tolerance for nearly 2.1 million models, or nearly 10\% of the entire models grid and we tag these models as class 'B' (for balanced). Does this mean that $\sim$ 90\% of our models are unphysical ? There are several reasons why this may not be the case. 

For more than 12 million models (or $\sim$60\% of the whole models grid) we have that  $L_{\star MIR}$ > 1.5$\cdot$(L$_{C, T_D}$-L$_{C, 12K}$), and we tag these models as class 'UB$_{M}$' (unbalanced, with mid-IR excess). There are several mechanisms that can make only a fraction of the input heating radiation effective to heat the intra-clump dust: 
\begin{itemize}
\item i) a significantly non-spherical clump geometry would justify lines of sight with drastically different optical depth, along which a major fraction of $L_{\star MIR}$ could almost freely escape absorption from the intra-clump dust;
\item ii) a similar effect would be produced by a spatial distribution of the embedded YSOs that is not particularly concentrated in the highest density regions of the clump; in case of YSOs located closer to the boundaries of the clump than to its center, the radiation emitted toward the near clump edge would travel a shorter column through lower density material and hence with a lower chance of being absorbed by clump dust, while the radiation emitted toward the far clump edge would be mainly absorbed in the clump central regions with a lesser chance to heat the far side of the clump. The result would be that $L_{\star MIR}$ would less effective in heating the clump dust; 
\item iii) a very clumpy medium with low clump filling factor would offer relatively lower optical depth lines of sight for the YSOs radiation to escape without significantly heating the clump dust; again, only a fraction of $L_{\star MIR}$ would be effective to heat the intra-clump dust.
\end{itemize}

Conversely we find nearly 5.9 million models (or about 30\% of the models grid) for which $L_{\star MIR}$ <0.6$\cdot$(L$_{C, T_D}$-L$_{C, 12K}$), which we tag as class 'UB$_{F}$' (unbalanced, with far-IR excess). An obvious mechanism that can heat the clump to temperatures higher than what could be justified by internal energy sources is irradiation from the ambient ISRF or from nearby bright objects like young OB associations. For example, a model with a very low fraction of forming YSOs and a \tdust =30K may not be unphysical at all as a nearby \hii\ region could easily heat the clump dust well above 30K (e.g. \citealt{Rollig+2011}) irrespectively of internal heating sources. Furthermore, the efficiency of such heating could be greatly increased in case of a clumpy medium where the low filling factor would allow ionising radiation to reach the inner regions of the clumps much more effectively (e.g. \citealt{Boisse1990}).

A more precise account of the effects of low filling factor in the clumps medium, different clump geometries and different spatial distributions of YSOs in the clump, and different ambient ISRF intensities would require detailed radiative transfer over a wide parameters space. This is a task that we reserve to carry out in a future paper for a very limited set of extreme cases, but that cannot certainly be offered for a grid of models as extensive as this one. For the scope of the present paper we have provided a number of  plausible qualitative arguments that support the physical plausibility of the large majority of the models in the grid. The models grid will contain the information about the mid/far-IR luminosity ratio so that a user will be alerted to specific conditions that likely need to be met for a certain model to be plausible, and will be able to verify if such conditions are present for any real source under study.

\section{New L/M evolutionary tracks}
\label{tracks}

\begin{figure*}
\centering
\includegraphics[width=0.49\textwidth]{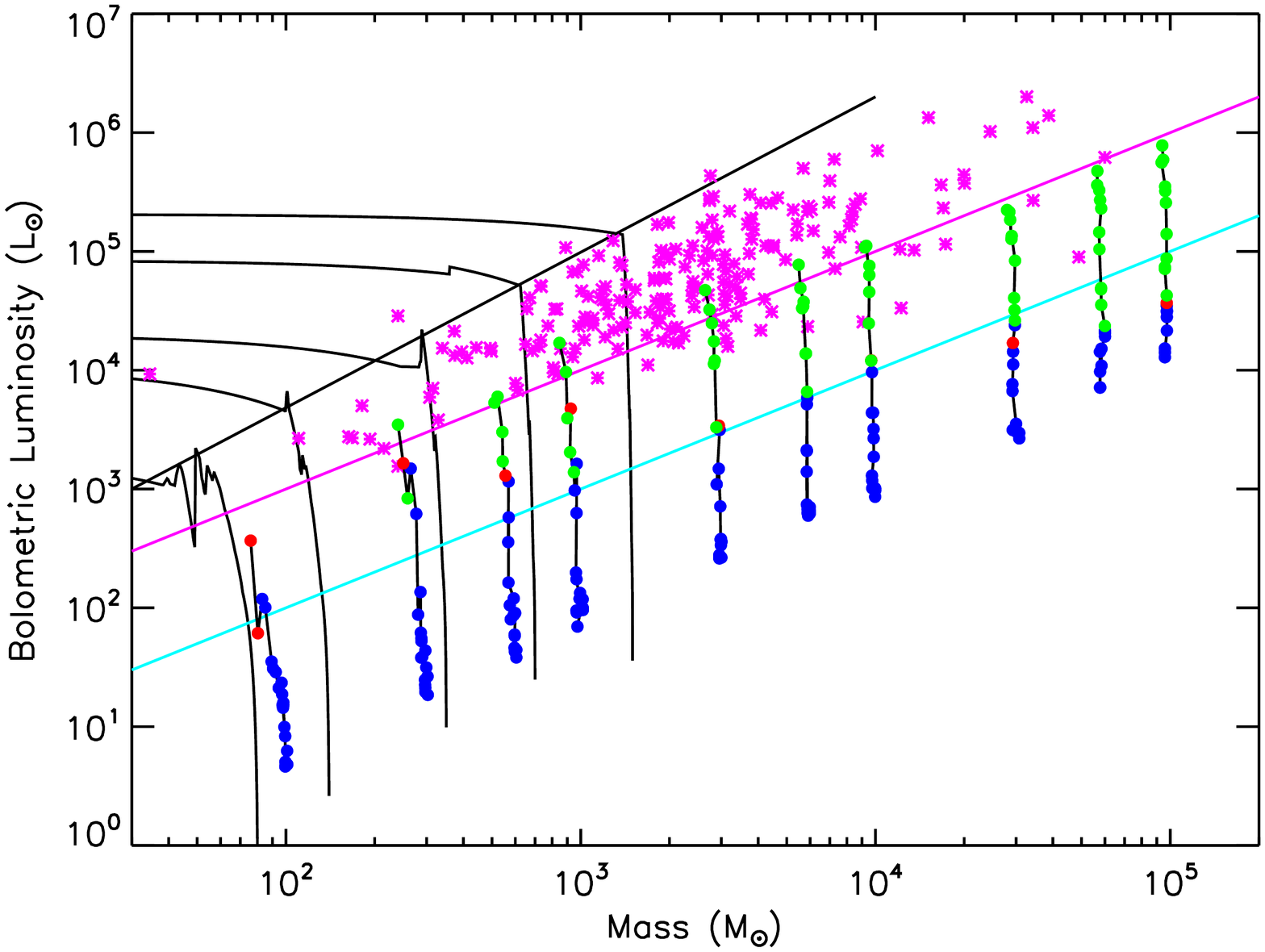}
\includegraphics[width=0.49\textwidth]{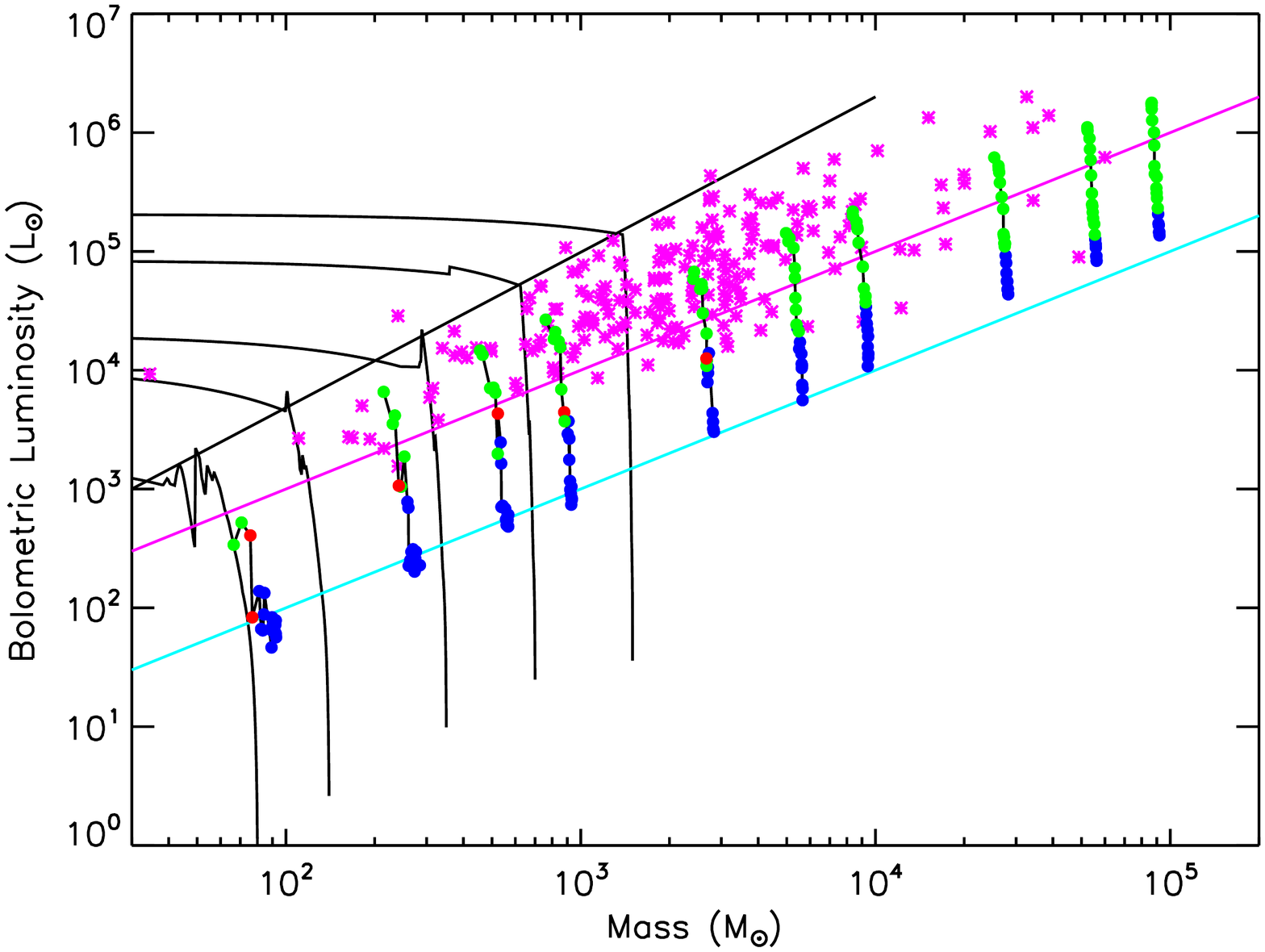} \\
\includegraphics[width=0.49\textwidth]{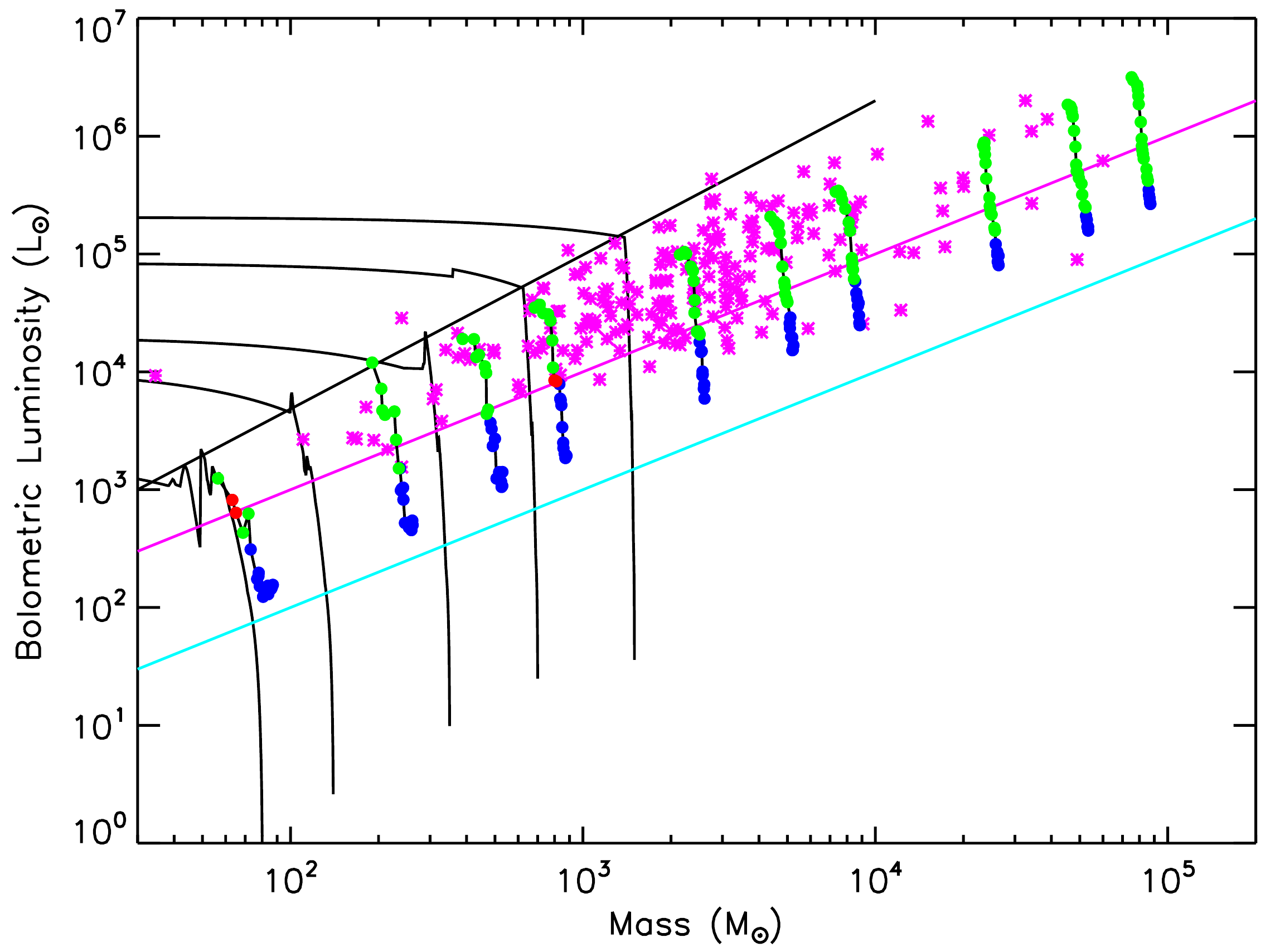} 
\includegraphics[width=0.49\textwidth]{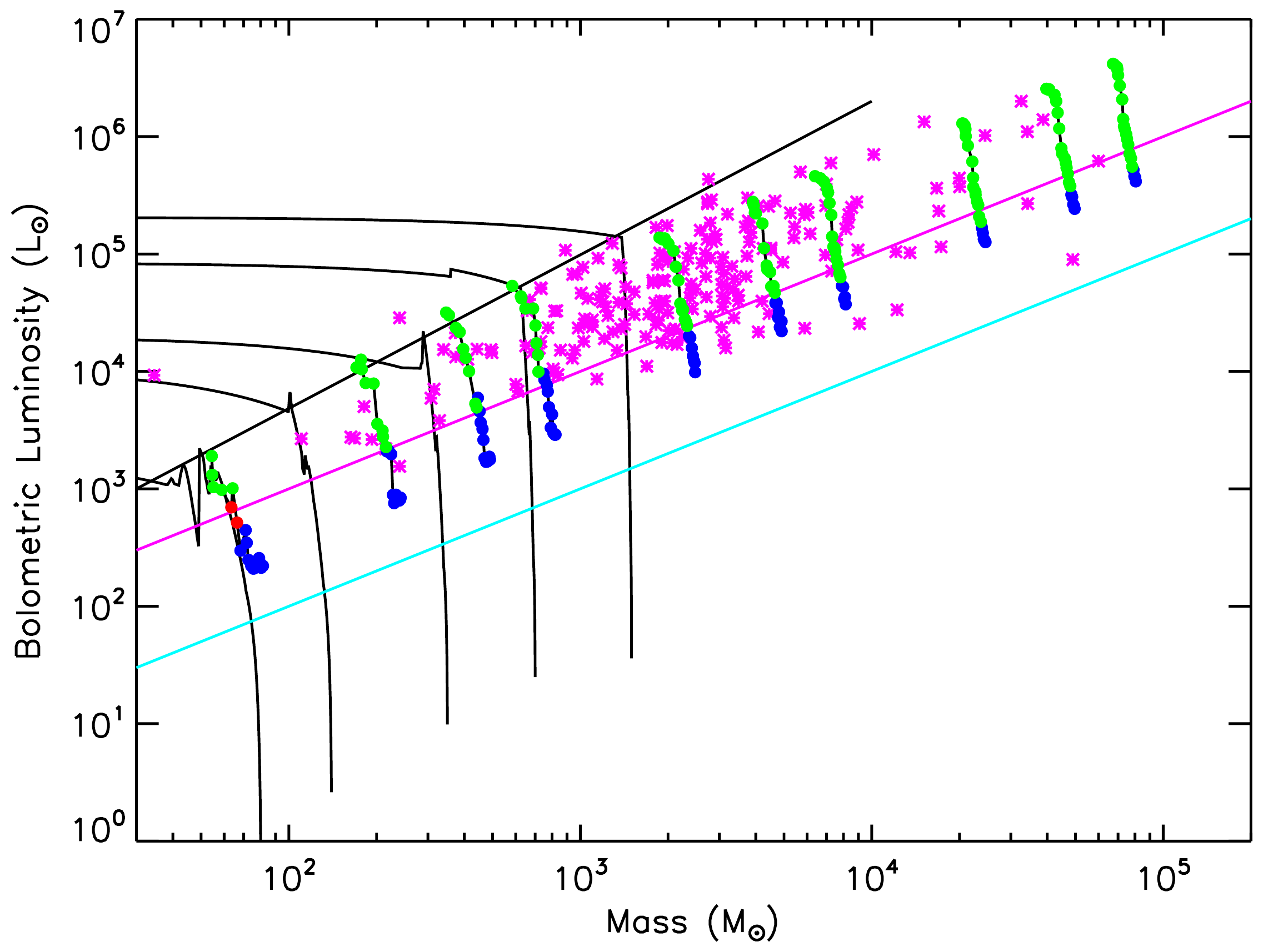} \\
\includegraphics[width=0.49\textwidth]{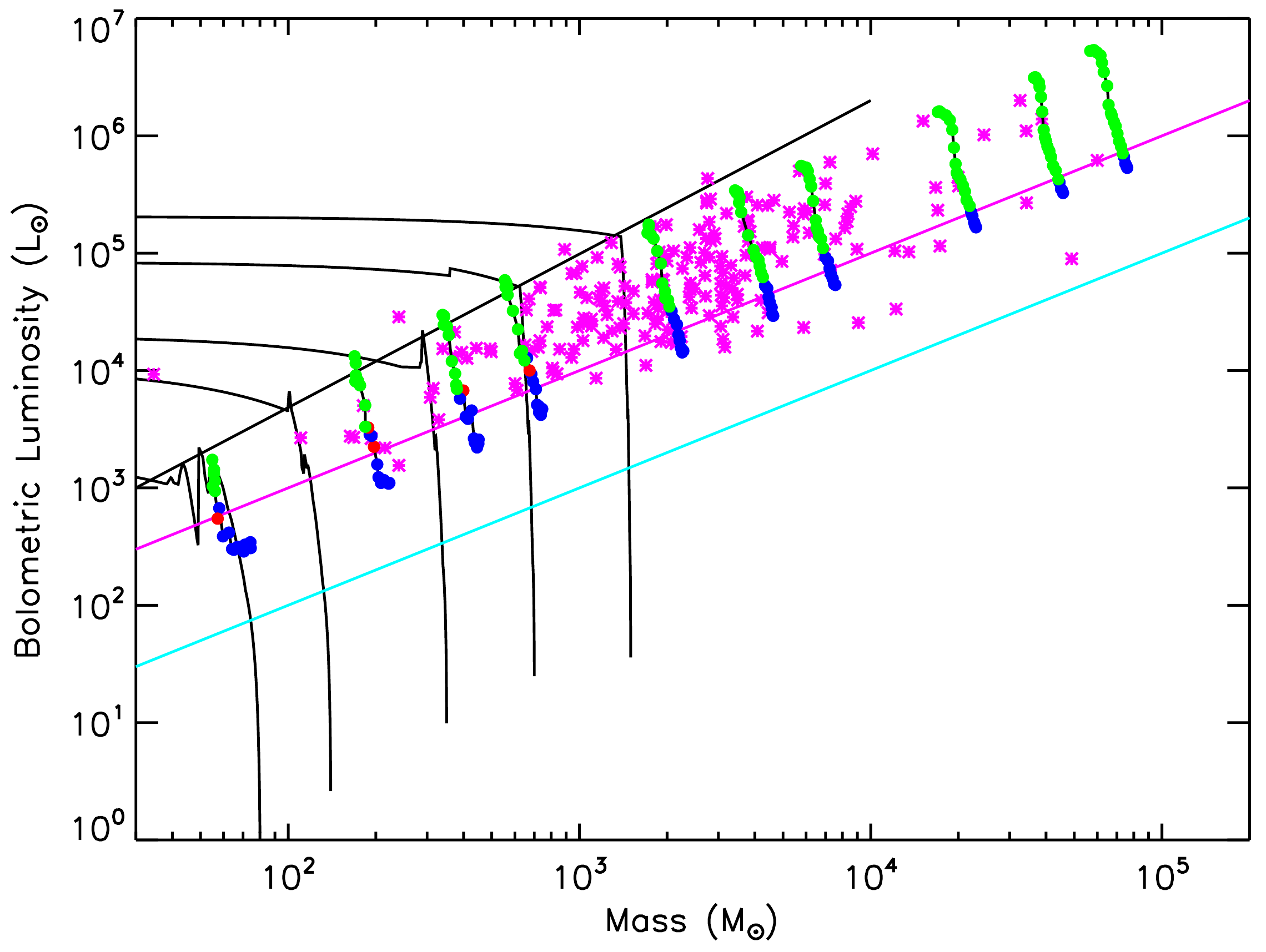}
\caption{\label{newtracks} L/M plots containing 10 new evolutionary tracks (the black lines connecting colored dots) built from the extended model grid as explained in the text; the luminosity is integrated under the models SEDs, and the mass is that of clump after excluding the fraction that at any time goes into stars (see text). The 5 full black lines are the old tracks from \protect\cite{Molinari+2008}, as in Fig. \ref{lm2}; cyan and magenta lines are also from Fig. \ref{lm2}. The five panels are for five different values of initial \fcore=[0.01, 0.1, 0.2, 0.3, 0.4] from left to right and top to bottom. Tracks are presented for 10 initial clump masses and for the entire clump age span 10$^4$-5$\times 10^5$ years. Symbol colours are coded on the ZAMS content of the model cluster; blue for models hosting no ZAMS stars, red for models hosting at least one ZAMS star, and green for models hosting at least one ZAMS star of spectral class B1.0. The magenta asterisks are the Hi-GAL clumps associated with UCHII from \protect\cite{Cesaroni+2015}.}
\end{figure*}

As already mentioned in \S\ref{modelsgrid} each model in the grid is created for a specific parameters set and replicated for ten Monte Carlo realizations. There is no attempt to consistently follow a specific model in time, as this would require to independently follow the evolution of the individual cores in each cluster model, which is beyond the scopes of the present work. As we are interested in a statistically significant characterization of a protocluster evolution, we will then build sequences of temporal evolution by sampling the grid of models following the evolution of the basic set of parameters. In this first attempt we will keep constant the intraclump dust temperature, although it could be expected that the intraclump dust will get warmed as the stellar content in the cluster evolves with time and stellar masses increase due to accretion from core and clump material. In this first realization we will let the \fcore\ parameter vary during the evolution, to reflect the global collapse of clump material that is feeding the cores as they evolve; in other words we follow competitive accretion protocluster models where the material that will eventually form the stars comes from regions not initially close or bound to the high density core fragments that constitute the initial seeds for collapse. As a result the initial population of protostars can accrete more efficiently with a larger supply of clump material, and a larger number of stars can form as more material flows into high-density and gravitationally unstable seeds \citep{Bonnell+2001}. Such large scale motions have been observed in systems as early as IRDCs and with rates that can reach 10$^{-3}$-10$^{-2}$\msunyr (e.g., \citealt{Peretto+2013, Kirk+13}). 

We then create tracks for 20 equally Log-spaced clump age steps from 10$^4$ to 5$\times 10^5$ years and for 10 representative initial clump masses of \mclump=[10$^2$, 3$\times 10^2$, 6$\times 10^2$, 10$^3$, 3$\times 10^3$, 6$\times 10^3$, 10$^4$, 3$\times 10^4$, 6$\times 10^4$, 10$^5$] \msun. We do five realization of such tracks for different initial values of clump mass locked in high-density cores \fcore = [0.01, 0.1, 0.2, 0.3, 0.4]. For each \fcore\ run we are letting this parameter increase from its initial value due to accretion from clump material not initially locked in cores; to estimate this increase we are assuming a steady clump$\leftrightarrow$cores accretion rate which is a function of clump mass and that can go from 5$\times 10^{-5}$ to 10$^{-2}$ \msunyr\ over the whole explored clump mass range (e.g., \citealt{He+2015, Tan+2014, Traficante+2017}). At each age step we compute the fraction of mass that is accreted onto dense cores and recompute the instantaneous \fcore\ value for the next age step. Likewise, at any given clump age we remove from the clump mass the fraction that has turned into stars and will no longer be traced by submillimeter continuum; this is needed to allow meaningful comparison with observations (e.g. Fig. \ref{newtracks}. With this new set of parameters we select the appropriate models and compute the median bolometric luminosity of the selected models to account for SED variance due to the other model parameters. At each age step we verify if the models contain ZAMS stars, and if at least one of these is a B1.0 spectral class or earlier. We then obtain the tracks reported in Fig. \ref{newtracks}, overlaid onto the old single-protostar clump evolutionary tracks of \cite{Molinari+2008} and the L/M thresholds determined by \cite{Molinari+2016c} based on the \ch\ gas temperature.

Let's first compare the new tracks with the old ones that were generated under the unrealistic hypothesis of a single massive star forming in the clump. A first thing is that in no case the new models can reproduce the full vertical rise of the old tracks toward what we called the "ZAMS line" in \cite{Molinari+2008} during the maximum age span of 5$\times 10^5$ years covered by the present models. The largest luminosity rise (in Log) is seen with models with initial \fcore=0.01 that, however, cannot reach beyond L/M$\sim10$. There is a clear overlap in the luminosity range explored for any given initial clump mass by tracks with different initial \fcore; for example the tracks for \fcore=0.1 start well below the endpoint of the \fcore=0.01 tracks. This is partially due to the overrun of \fcore\ caused by the fact that this parameter is free to grow in time due to clump-to-cores accretion; indeed we verify that for any given initial \mclump, each model track will end its path with a final \fcore\ that is larger than the initial value of the model for the initial \fcore\ immediately above. Let's take for example the models for initial \mclump=1000\msun, that is the third track form the left in each panel of Fig. \ref{newtracks}. The model with initial \fcore=0.01 (panel a) will have \fcore=0.18 at the top of the track, which is larger than the initial \fcore\ at the start of the analogous track in panel b; this is generally the case for all tracks. However, we verified that when for the same initial \mclump\ a track for a given initial \fcore\ will reach the luminosity similar to the start of the subsequent \fcore\ track, its instantaneous \fcore\ value will always be lower. This means that there is also an age effect as older clusters will have had more time to accrete cores mass into stars even for a smaller value of \fcore. In other words although the tracks in Fig. \ref{newtracks}a will have a lower \fcore\ compared to the portions of the tracks in \ref{newtracks}b in the luminosity range where they overlap, the actual stellar mass is higher in the former case and as a result the bolometric luminosity will be higher. This effect, present also to a variable degree for the other \fcore\ tracks sets, was anticipated in Figs. \ref{zamsfrac} and \ref{zamsnozams} and it is verified here in detail.

A second aspect is the inclination of the new tracks as they ascend toward higher luminosity. In the present models more mass is lost from the clump as the formation of a cluster produces much more stellar mass than a single forming object. As an example the final total stellar mass for an initial \mcl=2000 \msun\ and \fcore=0.2 amounts to nearly 600\msun, or about 17 times more stellar mass than in the case of a single forming object. As a result, the mass lost from the clump gas is much higher and that explains the tracks "leaning" to the left.

A third and most important aspect is that protostars in these evolving protoclusters reach the ZAMS earlier than the old tracks would suggest. The "ZAMS line" we drew for our 2008 tracks was matched to the location of UCH{\sc ii} regions whose luminosity was mostly based on IRAS fluxes, and was found to nicely match with the model single forming stars reaching stellar masses and radii typical of ZAMS stars. In the present case the stellar mass is distributed along with an IMF, resulting in lower mass value for the highest mass star, that will then evolve slower. The 2000\msun\ track of \cite{Molinari+2008} was covered in $\sim$1.5$\times 10^5$ years, while the present tracks are covered in 5$\times 10^5$ years. Intermediate mass ZAMS stars (red symbols in Fig. \ref{newtracks}) start to appear much earlier than the old tracks was suggesting, and with a mass that of course depends on the clump mass. Virtually all tracks in their initial age steps do not have ZAMS stars inside, but they appear faster the higher is \mclump\ and the higher is \fcore. The appearance of higher mass ZAMS (e.g., earlier that B1, or stellar mass higher than $\sim$12\msun) will happen at \sftime$\gsim10^5$ years for \fcore=0.01 (Fig. \ref{newtracks}a) and, as there is a lower stellar mass compared to higher clump masses and \fcore, at lower bolometric luminosities. For higher \fcore\ and \mcl\ the IMF sampling statistics makes it more likely that objects earlier than B1.0  will be found soon (Fig. \ref{newtracks}c,d,e for the higher \mcl). 

It is then clear that our notion of a "ZAMS line"  in the L/M diagram no longer applies when considering the formation of stars in protoclusters. Instead, we introduce the notion of a "ZAMS strip" whose exact location depends on the models parameters; for \fcore $\geq0.2$ such strip would correspond to the portion of the plane where \loverm $\geq$10 that \cite{Molinari+2016c} identified as a threshold where the dense gas temperature traced by \ch\ rotational lines starts to increase with L/M, while for lower L/M the temperature was not varying or the \ch\ transitions were not even detected. The proposed threshold is also found in excellent agreement with the distribution of most of the Hi-GAL dense clumps that \cite{Cesaroni+2015} find associated with UCH{\sc ii} from CORNISH \citep{Purcell2013}. A few of these dense clumps with radio counterpart with L/M$\leq$ 10 seem compatible with models with \fcore$\leq$0.1 that, especially for \mcl$\geq 10^4$\msun, see the presence of B1.0 ZAMS stars or earlier.

As a final cautionary remark, we point out that in our models we have neglected the effect of feedback from the forming YSOs back into the clump gas. Molecular outflows and radiation from newborn stars will inject energy and momentum in the clump gas, possibly slowing down the star formation process or truncating the YSOs mass distribution once high-mass stars will have formed, with modes and efficiencies that will also depend on the spatial distribution of clump gas and embedded YSOs.

\section{Applications}
\label{applications}

To test the predictive power of our grid of models on real SED data we follow two approaches. In the first we use data from the Spitzer and Herschel surveys of the Orion A star forming region, where single cores are resolved with these facilities \citep{Polychroni+2013}; in this way we can test our models grid in a controlled environment where we know the properties (e.g., evolutionary stage, mass) of the underlying YSO population. However, the Orion A star forming region is known to form low and intermediate mass YSOs; to extend the testing of the models grid to high-mass star formation we also attempt fit to SEDs of massive dense clumps hosting high-mass YSOs.

\subsection{SED fitting of Orion A simulated clumps}
\label{sim_clumps}

The approach we followed is to create simulated clump SEDs at various distances, using mapping data from Herschel in the Far-Infrared. Herschel maps of Orion A were processed following the method illustrated by \cite{Baldeschi+2017a} to simulate the appearance of this region at distances of 2, 3, 5 and 7 kpc if observed with the Herschel cameras. Source extraction, photometry and band-merging SED building were carried out in an automatic way following methodology outlined in \cite{Elia+2017}. To complete the simulated clumps SEDs with mid-IR data, we used the compilation of Spitzer data for the YSOs in the HOPS Herschel program from \cite{Furlan+2016} and, for each IRAC band and the MIPS 24\um\ band, we co-added the fluxes of all the sources falling within the FWHM size at 250\um\ of each clump detected in the `receded' maps, for all simulated distances. We obtain 33, 21, 9  and 8 clumps with a complete 3-500\um\ SED detected in the Herschel receded maps at 2, 3, 5 and 7kpc respectively, that were fitted to our new models grid. 

The fitting is carried out as a normal $\chi ^2$ minimization, in which statistical weights are distributed among the data points so that the minimization is not driven by one or two data points with very low uncertainty compared to the rest of the data in the SED. To do this, we divide the full spectral range in three sections: $\lambda \leq 25$\um, 70\um$\leq \lambda \leq$250\um, and $\lambda\geq 350$\um; equal weights are given to the three ranges, and uncertainties to each data point are modified accordingly. This procedure has the additional advantage that an SED is consistently fitted over the entire wavelength range irrespectively of the more or less dense data coverage in each of the three ranges. 
We select all models where $\chi ^2 \leq$ min($\chi ^2$)+3 and fit the distribution of modelled clump ages \sftime, and mass fraction in dense cores \fcore\ with a Gaussian to determine the mode of the distribution and get an estimate for the spread of the parameters. This is somewhat artificial because there is no reason why the distribution of the parameters for the fitted models should obey a normla distribution; there is also little scope for a more detailed statistical analysis because the models that can fit an observed SED may belong to very different areas of the parameterts space. Here we just want to get an idea of what is, if any, a dominant trend that emerges. For a number of simulated clumps (9 at 2kpc, 1 at 3kpc and 3 at 5kpc) the fit to the models were not selective for some of the parameters (typically the clump age), and were not considered in the subsequent analysis.

\begin{figure*}
\begin{center}
\includegraphics[width=\textwidth]{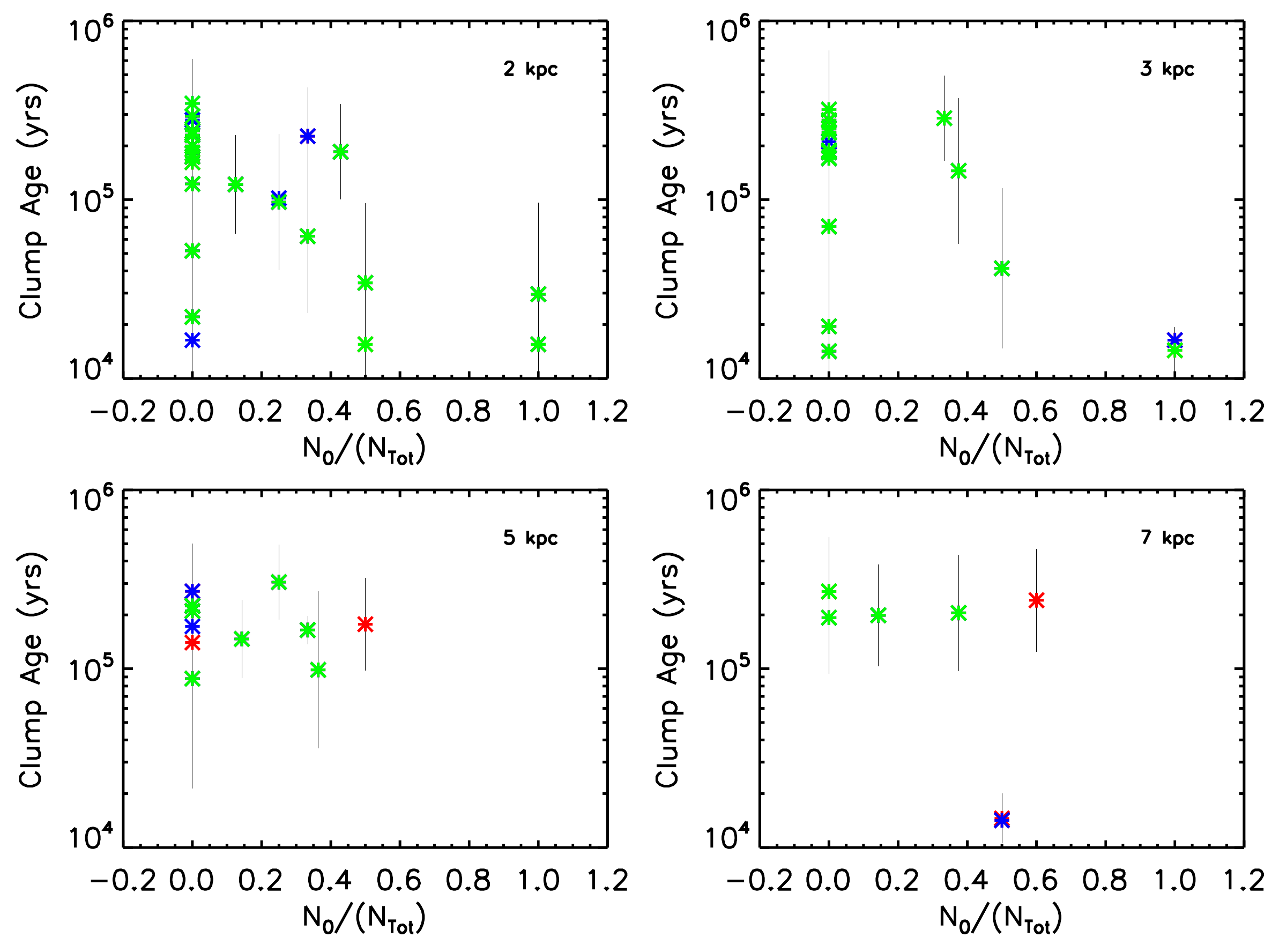}
\caption{Clump age estimated from SED fitting to our models grid as a function of the fractional number of Class 0 YSOs associated with each simulated clump (from \citealt{Furlan+2016}) for simulated Orion A clumps at the distance indicated in the panels. Points are color-coded depending on model class (see \S\ref{balance}): green for B, blue for UB$_M$ and red for UB$_F$ models. }
\label{age_n0}
\end{center}
\end{figure*}

Figure \ref{age_n0} shows the most likely age of the simulated Orion A clumps as determined from the fit to our models grid, as a function of the fractional number of Class 0 YSOs associated with each clump for each simulated distance. The evolutionary nature and properties of the underlying population of YSOs is taken from the compilation of \cite{Furlan+2016} for the HOPS project, who classify YSOs as Class 0, I, II and `flat', based on the 8-24\um\ SED shape. The figures show a trend of decreasing fitted age with increasing fraction of Class 0 YSOs found associated with the clumps for simulated distances of 2 and 3 kpc, confirming the ability of our models to diagnose global trends in clump ages. This indication is absent in the results for the larger simulated distance; we believe this is due to the fact that at larger distances the linear size of the simulated clumps also increases, encompassing a relatively higher number of more evolved YSOs that tend to be more sparsely distributed in the OriA region compared, to Class 0-I objects. This is also confirmed by the fact that the fractional number of Class 0 YSOs in the OriA clumps never reaches 1 for simulated distances of 5 and 7 kpc. It is therefore plausible that larger clumps (those at higher distances) tend to be fitted with models of relatively higher ages.

\begin{figure*}
\begin{center}
\includegraphics[width=\textwidth]{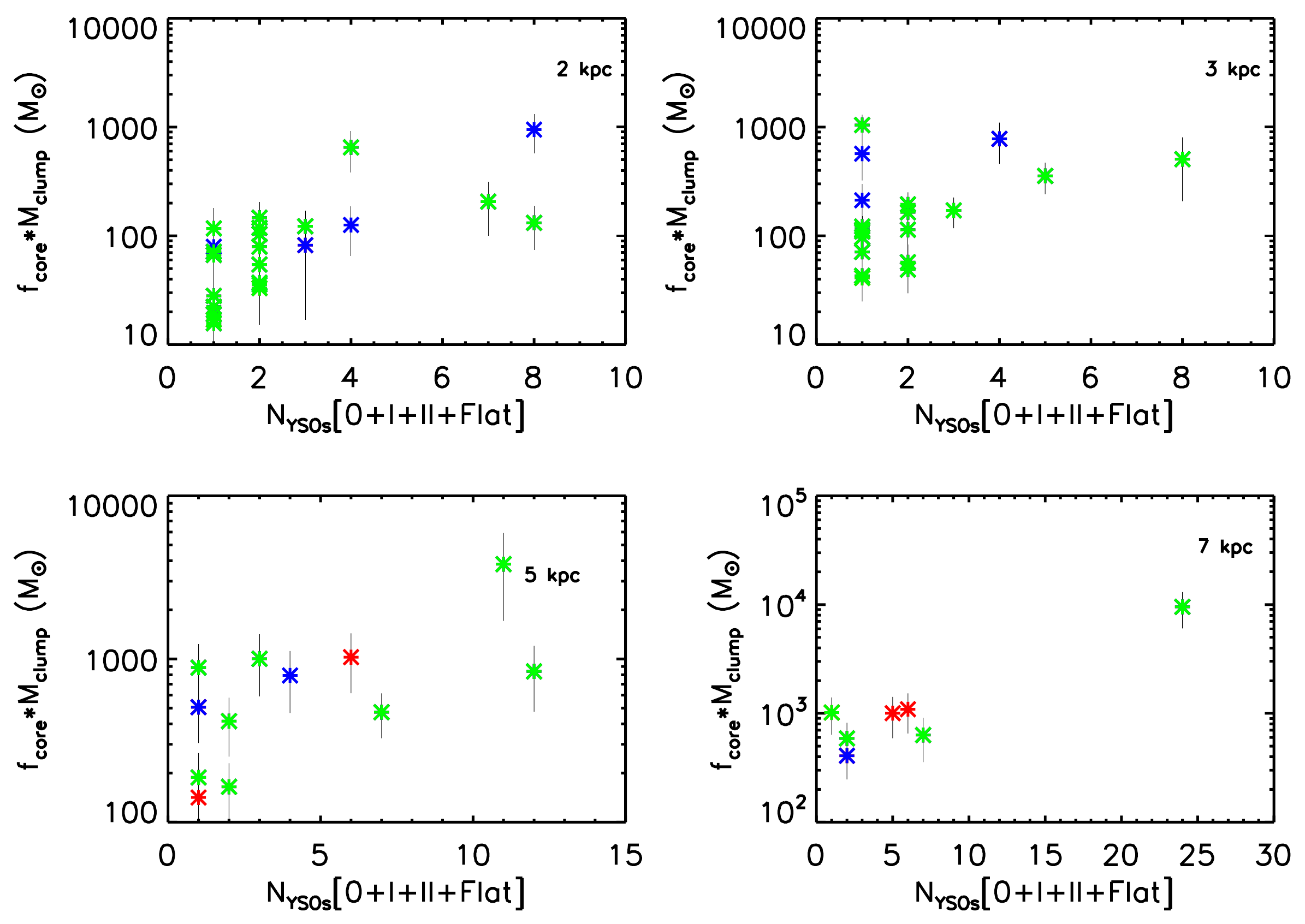}
\caption{Simulated clump fractional mass in dense cores \fcore x \mclump\ estimated from SED fitting to our models grid as a function of the total number of YSOs (class 0, I, II and Flat spectrum) associated with each simulated clump (from \citealt{Furlan+2016}) for  Orion A clumps simulated at the distance indicated in the panels. Color coding of the points is as in Fig. \ref{age_n0}.}
\label{fcore_nyso}
\end{center}
\end{figure*}

Another check we can do with the fit to the simulated Orion A clumps is the fraction of mass locked in dense cores forming YSOs. Figure \ref{fcore_nyso} reports the relationship between the cores fractional mass as determined form the models fit, as a function of the total number of YSOs of all evolutionary classes from \cite{Furlan+2016} that are positionally within the FWHM 250\um\ size of each simulated clump. Although \cite{Furlan+2016} also provide masses for their protostar+disk+envelope YSOs, they would be of limited use in the present context as these authors assume fixed protostellar mass of 0.5\msun\ independently of evolutionary stage and  luminosity and a fixed disk mass of 0.05\msun; only the envelope mass is determined from a model fit to the YSOs SEDs and is generally a small fraction of total YSOs mass.
The positive trend between the mass locked in star-forming dense cores as modelled in our simulated clumps and the number of associated YSOs is a  positive qualitative indication that our models are able to track in a relative way the young stars yield of a clump.

\subsection{Two examples of dense massive clumps}
\label{massive_clumps}

We present in this section two examples of the fitting of observed SEDs to the models grid for two different compact clumps from the Hi-GAL survey. The SEDs are assembled using all available photometry from the near-infrared to the millimeter (WISE, Hi-GAL, ATLASGAL, BGPS), and the contributions from multiple resolved sources below 160\um\ are coadded in order to produce a clump-level integrated SED. 

\begin{figure*}
\begin{center}
\includegraphics[width=0.49\textwidth]{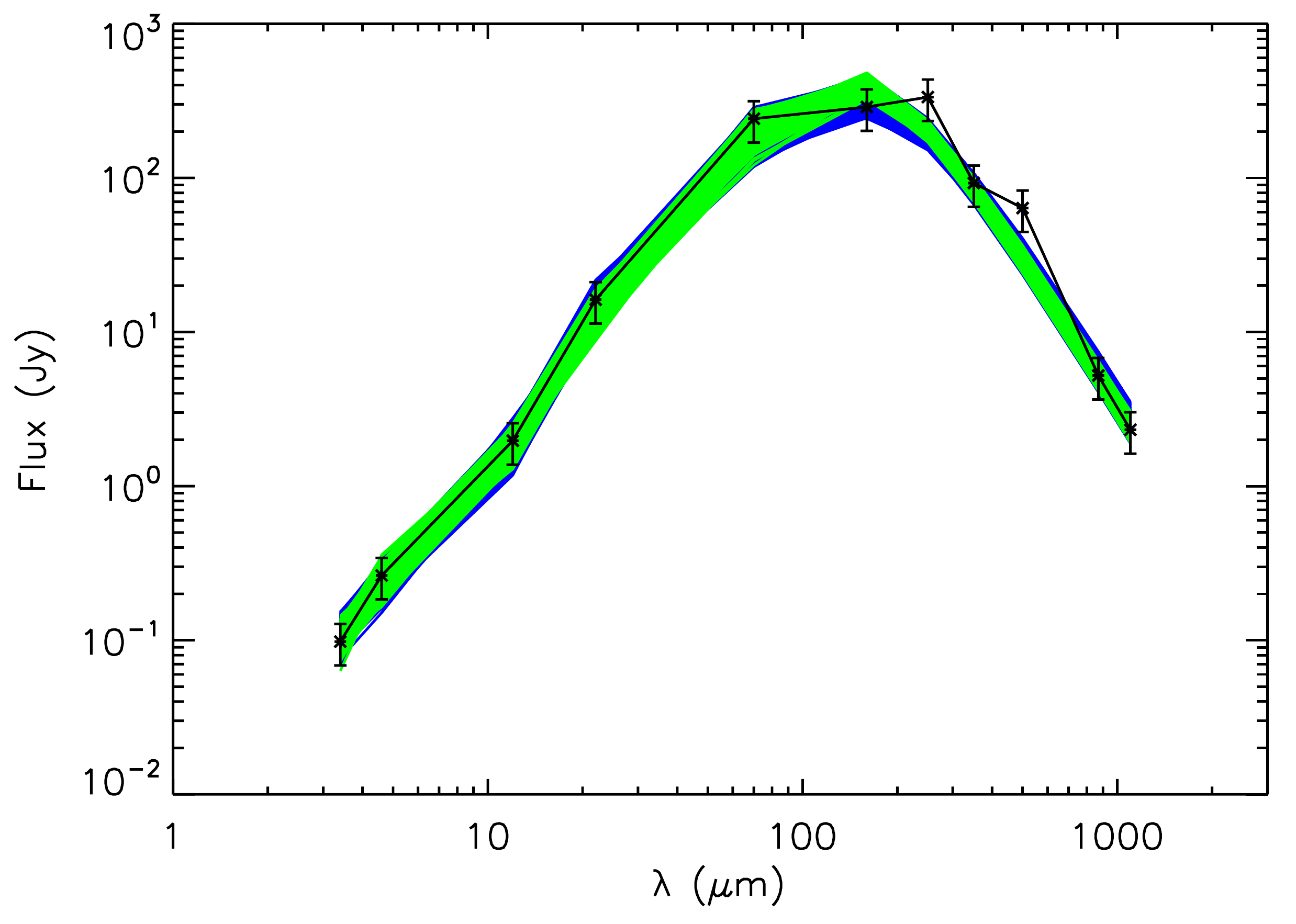}
\includegraphics[width=0.49\textwidth]{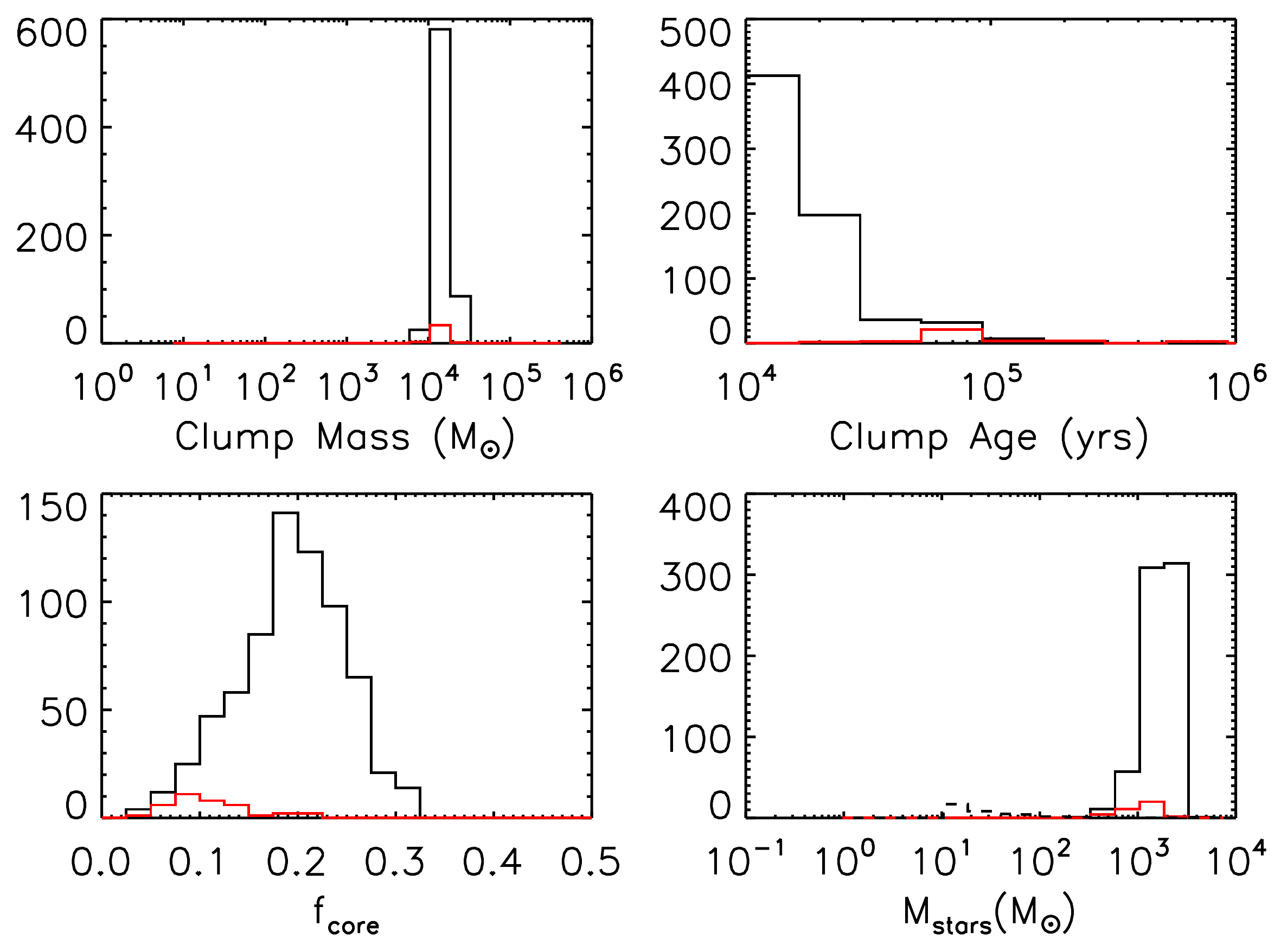}
\caption{\textit{a (left):} Observed clump-integrated SED of IRAS19092+0841 (black line connecting asterisks) with superimposed the best fit models with $\chi^2 \leq min(\chi^2)+3$. SEDs are color coded as in Fig. \ref{age_n0}. \textit{b (right):} Distribution of \mclump, age, \fcore\ and total stellar mass for the set of models shown in the left panel; models containing ZAMS stars are identified with the red histograms in the right panel. The dashed histogram in the bottom-right corner of the right panel reports the distribution of total mass of ZAMS stars present in the protocluster.}
\label{fig_mol98}
\end{center}
\end{figure*}

\subsubsection{IRAS19092+0841: a relatively young clump}

\begin{figure*}
\begin{center}
\includegraphics[width=0.49\textwidth]{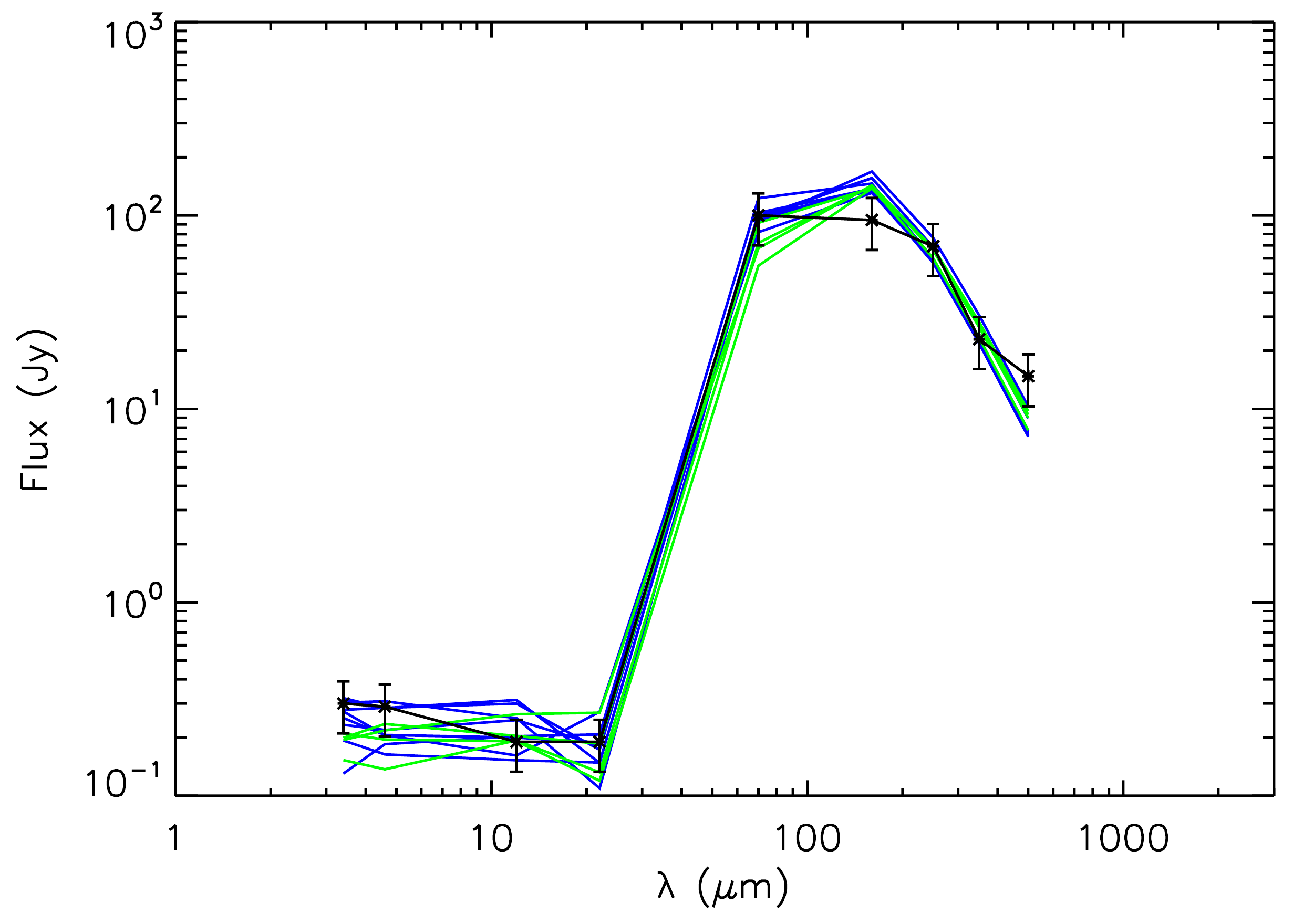}
\includegraphics[width=0.49\textwidth]{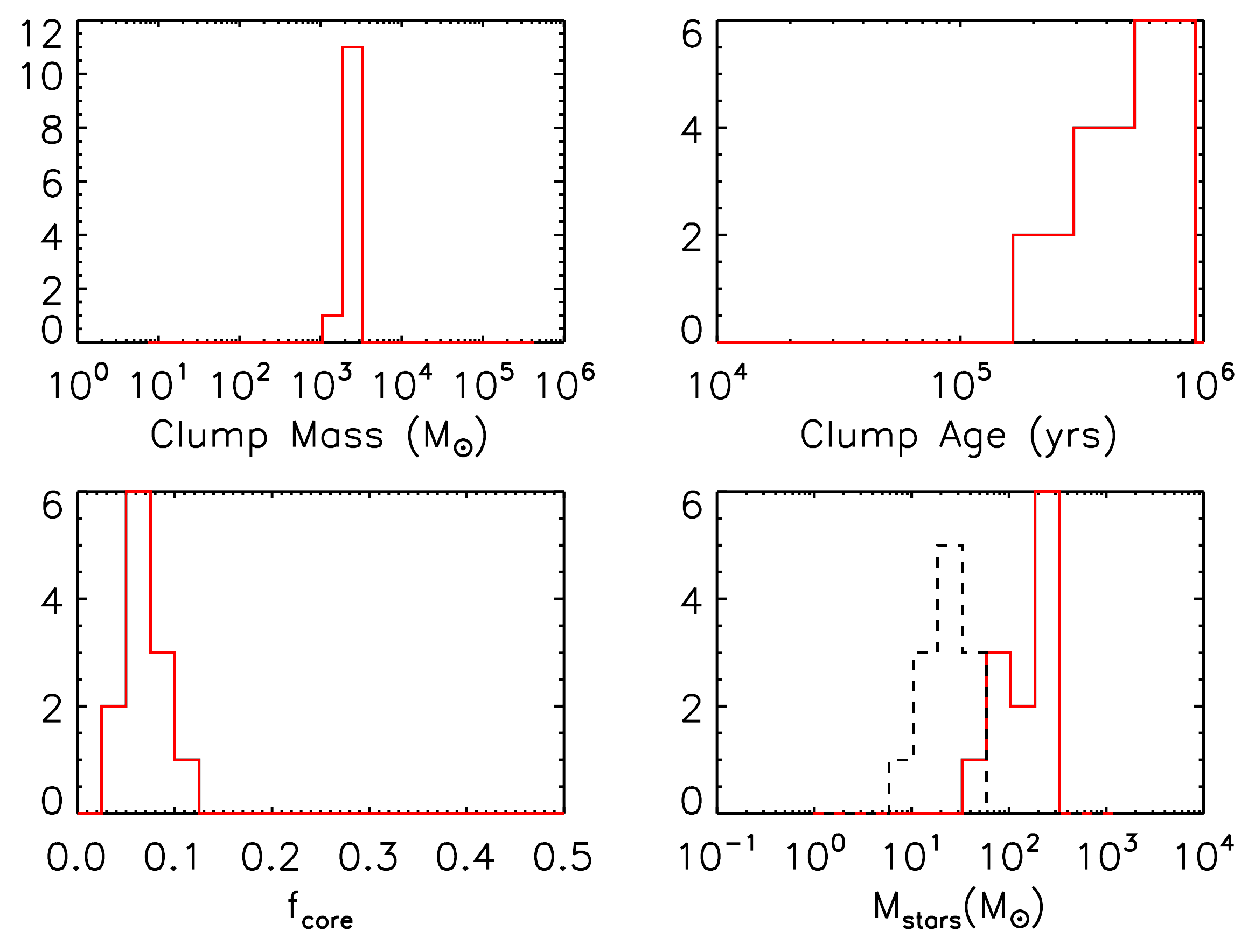}
\caption{Same as Fig. \ref{fig_mol98} for IRAS05137+3919. }
\label{fig_mol8}
\end{center}
\end{figure*}

IRAS19092+0841 is catalogued as source \#98 in the sample of \cite{Molinari+1996}; it is located at [l, b]=[43.0379, -0.4531] at a distance of 8.3kpc \citep{Elia+2017}. The source consists of a bright and compact far-IR/submm clump that is resolved in at least 4 counterparts at $\lambda \leq$24\um. The IRAS colours differ from those typical of UCH{\sc ii} regions and as such was classified as "\textit{low}" in \cite{Molinari+1996}; the proposed early evolutionary stage was confirmed by the absence of any radio continuum counterpart at 5-15GHz at a level of 0.1 mJy/beam. Figure \ref{fig_mol98}a shows the observed fluxes of the source from 3.41\um\ to 1.1mm from WISE, Hi-GAL, ATLASGAL and BGPS; the fluxes of all WISE counterparts contained in the area corresponding to the source size at 250\um\ were coadded. The SED fitting to the models grid identifies 693 models where $\chi^2 \leq min(\chi^2)+3$, and the corresponding distribution of the physical parameters is shown in Fig. \ref{fig_mol98}b. The clump mass is estimated $\sim 1.4\times 10^4$ \msun\ with a favoured clump age of $\sim 1.5\times 10^4$ yrs and a temperature for the intra-clump dust T$\sim$20K; only 37 out of 693 models contain at least 1 ZAMS star, and the models suggest a fraction of mass locked in cores of the order of 20\%. The 693 best fit models (in the $\chi^2$ range selected) are all very similar in shape, and isolate a relatively well  defined area of parameters space. 

The models  suggest that a cluster of young protostars is present, with a current total stellar mass of $\sim$1800\msun, corresponding to a present-day Star Formation Rate of 0.13 \msun/yr. Recently \cite{Veneziani+2017} provided a method to estimate the Star Formation Rate of massive clumps from their mass. In short, upgrading an earlier prescription from \cite{Veneziani+2013}, the clump mass is compared to our earlier set of evolutionary tracks \citet{Molinari+2008} where the clump, instead of a single forming massive star, hosts a cluster whose mass is simulated using Monte Carlo realizations of a star cluster following the IMF of \cite{Kroupa2001} and producing the same bolometric luminosity typically observed for clumps of the same total mass but associated with UCH{\sc ii} regions. The SFR computed according to this method in IRAS19092+0841 is $\sim 6.3\times 10^{-4}$\msunyr, that is about 200 times lower than the value given by the present tracks (0.13, see above). This is not surprising as in this work we are comparing clumps masses and luminosities with tracks that provide the present estimated embedded stellar mass and age, while \cite{Veneziani+2017} uses simulated embedded stellar masses and ages at the time the clump has reached luminosities typical of UCH{\sc ii}, therefore providing a terminal estimate of the SFR of the clump averaged over its early evolutionary path. Quite speculatively, the comparison would suggest that in IRAS19092+0841 we are probably looking at a very recent burst of star formation, where the future evolution will see an increase in the luminosity of the present protostellar objects rather than a further increase in their number.

\subsubsection{IRAS05137+3919: a relatively evolved clump}

IRAS05137+3919 is catalogued as source \#8 in the sample of \cite{Molinari+1996}; it located at [l, b]=[168.0626, 0.822] at a distance of 7.7kpc \citep{Elia+2019}. The source consists of a bright and compact far-IR/submm clump close to a much fainter companion about 30\arcsec\ to the south-east in the SCUBA map of \cite{Molinari+2008}; the brighter northern component was detected in interferometry at 3.4mm continuum with OVRO \citep{Molinari+2002}, and it is associated with 4 YSOs at WISE wavelength. The IRAS colours were  typical of UCH{\sc ii} regions and as such was classified as "\textit{high}" in \cite{Molinari+1996}; the proposed relatively late evolutionary stage was confirmed by the presence of a radio continuum counterpart at 8.3GHz with a peak flux of 1.2 mJy \citep{Molinari+2002}.

The SED fitting to the models grid identifies 12 models where $\chi^2 \leq min(\chi^2)+3$, and the corresponding distribution of the physical parameters is shown in Fig. \ref{fig_mol8}b. The clump mass is estimated close to 2000\msun\ with a favoured clump  age of $\sim 5.3\times 10^5$ yrs, and a temperature of the intra-clump dust T$\sim$24K; all of the models contain ZAMS stars, with a fraction of mass locked in cores $\sim$7\%. The 12 best fit models (in the $\chi^2$ range selected) are all very similar in shape, and isolate a well defined area of parameters space. The models suggest that a cluster of young protostars is present, with a total stellar mass of $\sim$200\msun, corresponding to a present-day Star Formation Rate of $3.8\times 10^{-4}$ \msun/yr, this time in relatively good agreement with the SFR value of $\sim 1.5\times 10^{-4}$ \msun/yr computed with the method of \cite{Veneziani+2017}. In this case our present tracks diagnose IRAS05137+3919 to be populated with ZAMS stars, representing a system that is closer (with respect to the previously examined source) to the terminal stage of its evolutionary path toward UCH{\sc ii} regions.

The mass of the most massive ZAMS object in the selected cluster models is in the range 8\msun $\leq M \leq 15$\msun; using standard assumptions as in \cite{bm98} these ZAMS stellar masses correspond to spectral classes between B2.3 and B0.5. The latter spectral class would also correspond to an emitted Lyman continuum of Log[N$_{Ly}$]=46.51 that agrees quite nicely with the value of 46.17 estimated directly from the radio continuum flux of the most intense radio source associated to the dense clump of IRAS05137+3919 as measured by \cite{Molinari+2002}.

\subsection{Models grid access}

The models presented in this paper are made available as a FITS file from the portal of the VIALACTEA Project\footnote{accessible at \url{http://vialactea.iaps.inaf.it}}. The VIALACTEA Project assembled photometric data and images for the Galactic Plane from a variety of continuum and spectroscopic surveys from the near-IR to the radio; on the VIALACTEA portal it is possible to access a Visual Analytics application that enables the access to photometric data from the near-IR to the radio of compact sources in the Galactic Plane, the construction of SEDs for compact sources and the fitting to the models grid. More details on the VIALACTEA features are described in \cite{Molinari+2017}, and on the VIALACTEA portal.

\section{Conclusions}
\label{conclusions}

We have created a new grid of 20 millions synthetic SED models for dense clumps hosting a forming cluster of YSOs. While SED models for individual YSOs that would typically form in a "core" are already available and widely tested using a variety of objects in nearby molecular clouds, the compact sources detected toward high-mass star forming regions and in general in Galactic Plane continuum far-IR and submillimeter surveys are systems with properties typical of "clumps" rather than "cores". As such, these structure typically harbour clusters of forming YSOs; no SED models were available for these class of objects. 

The fundamental parameters of the models that we presented in this work are the total mass of the dense clump, the temperature of the bulk of the dust in clumps, the fraction of clump material that is locked in dense cores, and the age of the clump expressed as the time since the start of the formation of the first YSO. Assuming a standard IMF and a uniform Star Formation History in the clump, these four parameters are sufficient to define the total mass and the upper age limit of the YSOs in each protocluster. Each YSO in these synthetic clusters is assumed to be a system consisting of a collapsing central object plus a circumstellar disk and a circumstellar envelope with a polar cavity. The SED of each YSO system in each simulated protocluster is sampled from the grid of models of R06 until the requested total mass in YSOs is reached. The YSOs are then statistically distributed inside the clump and the effect of intra-clump dust extinction is taken into account. Finally, we co-add all YSOs SEDs together with the SED from the intraclump dust emission to yield the integrated clump SED.

A number of parameters of the YSO star+disk+envelope system are not explicitly controlled in our simulations (YSO disk radius, envelope radius, disk inclination angle, amplitude of polar cavity angle), but we partially account for the variance that they can introduce in the resulting SED by carrying out 10 independent Monte Carlo realizations of the protocluster simulation for each combination of the 4 fundamental parameters that we use.

We illustrated the main features of the models grid, and discussed the effect of the different parameters choice on the resulting SEDs. Degeneracies in the parameters space are present, but if a sufficient number of SED photometric points are available and the distance to the objects is known the degeneracies can be substantially controlled. Tests on the Orion A star forming region and two high-mass dense clumps from the Hi-GAL survey show that best fitting models define a substantially confined area of the parameters space.

A new set of evolutionary tracks in the L/M plane has been presented, updating and extending the work of \cite{Molinari+2008} to more realistic systems where clusters rather single massive objects are forming in dense clumps; yet the general trends identified in \cite{Molinari+2008} are confirmed, reinforcing the robustness of the L/M diagram as a diagnostic tool for the evolutionary classification of dense star-forming clumps. 

The models fit tests presented on real objects confirm the models ability to diagnose the presence of newborn ZAMS stars, proposing simple colour-based diagnostics to identify young clumps potentially hosting such stars.

\section*{Acknowledgements}

It is a real pleasure to thank an anonymous referee for careful and insightful reports that helped to considerably improve the quality of the paper. This work has been funded by VIALACTEA, a Collaborative Project under Framework Programme 7 of the European Union (Contract \# 607380),  that is hereby acknowledged.




\bibliographystyle{mnras}
\bibliography{/Users/molinari/Science/sergio_bib} 

\begin{thebibliography}{}
\makeatletter
\relax
\def\mn@urlcharsother{\let\do\@makeother \do\$\do\&\do\#\do\^\do\_\do\%\do\~}
\def\mn@doi{\begingroup\mn@urlcharsother \@ifnextchar [ {\mn@doi@}
  {\mn@doi@[]}}
\def\mn@doi@[#1]#2{\def\@tempa{#1}\ifx\@tempa\@empty \href
  {http://dx.doi.org/#2} {doi:#2}\else \href {http://dx.doi.org/#2} {#1}\fi
  \endgroup}
\def\mn@eprint#1#2{\mn@eprint@#1:#2::\@nil}
\def\mn@eprint@arXiv#1{\href {http://arxiv.org/abs/#1} {{\tt arXiv:#1}}}
\def\mn@eprint@dblp#1{\href {http://dblp.uni-trier.de/rec/bibtex/#1.xml}
  {dblp:#1}}
\def\mn@eprint@#1:#2:#3:#4\@nil{\def\@tempa {#1}\def\@tempb {#2}\def\@tempc
  {#3}\ifx \@tempc \@empty \let \@tempc \@tempb \let \@tempb \@tempa \fi \ifx
  \@tempb \@empty \def\@tempb {arXiv}\fi \@ifundefined
  {mn@eprint@\@tempb}{\@tempb:\@tempc}{\expandafter \expandafter \csname
  mn@eprint@\@tempb\endcsname \expandafter{\@tempc}}}

\bibitem[\protect\citeauthoryear{{Aguirre} et~al.,}{{Aguirre}
  et~al.}{2011}]{Aguirre2011}
{Aguirre} J.~E.,  et~al., 2011, \mn@doi [\apjs] {10.1088/0067-0049/192/1/4},
  \href {http://adsabs.harvard.edu/abs/2011ApJS..192....4A} {192, 4}

\bibitem[\protect\citeauthoryear{{Baldeschi} et~al.,}{{Baldeschi}
  et~al.}{2017}]{Baldeschi+2017a}
{Baldeschi} A.,  et~al., 2017, \mnras, 466, 3682

\bibitem[\protect\citeauthoryear{{Benjamin} et~al.,}{{Benjamin}
  et~al.}{2003}]{Benjamin+2003}
{Benjamin} R.~A.,  et~al., 2003, \mn@doi [\pasp] {10.1086/376696}, \href
  {http://adsabs.harvard.edu/abs/2003PASP..115..953B} {115, 953}

\bibitem[\protect\citeauthoryear{{Bergin} \& {Tafalla}}{{Bergin} \&
  {Tafalla}}{2007}]{BT2007}
{Bergin} E.,  {Tafalla} M.,  2007, \araa, 45, 339

\bibitem[\protect\citeauthoryear{{Beuther} et~al.,}{{Beuther}
  et~al.}{2015}]{Beuther+2015}
{Beuther} H.,  et~al., 2015, \aap, 581, 119

\bibitem[\protect\citeauthoryear{{Binney} \& {Merrifield}}{{Binney} \&
  {Merrifield}}{1998}]{bm98}
{Binney} J.,  {Merrifield} M.,  1998, Galactic Astronomy.
Princeton Series in Astrophysics, Princeton University Press

\bibitem[\protect\citeauthoryear{{Boiss{\'e}}}{{Boiss{\'e}}}{1990}]{Boisse1990}
{Boiss{\'e}} P.,  1990, \aap, 228, 483

\bibitem[\protect\citeauthoryear{{Bonnell}, {Bate}, {Clarke}  \&
  {Pringle}}{{Bonnell} et~al.}{2001}]{Bonnell+2001}
{Bonnell} I.~A.,  {Bate} M.~R.,  {Clarke} C.~J.,   {Pringle} J.~E.,  2001,
  \mnras, 323, 785

\bibitem[\protect\citeauthoryear{{Carey}, {Noriega-Crespo}, {Mizuno}, {Shenoy},
  {Paladini}  \& {et al.}}{{Carey} et~al.}{2009}]{Carey+2009}
{Carey} S.~J.,  {Noriega-Crespo} A.,  {Mizuno} D.~R.,  {Shenoy} S.,  {Paladini}
  R.,   {et al.} 2009, PASP, 121, 76

\bibitem[\protect\citeauthoryear{{Cesaroni} et~al.,}{{Cesaroni}
  et~al.}{2015}]{Cesaroni+2015}
{Cesaroni} R.,  et~al., 2015, \aap, 579, 71

\bibitem[\protect\citeauthoryear{{Csengeri}, {Urquhart}, {Schuller}, {Motte},
  {Bontemps}, {Wyrowski}  \& {et al.}}{{Csengeri} et~al.}{2014}]{Csengeri+2014}
{Csengeri} T.,  {Urquhart} J.~S.,  {Schuller} F.,  {Motte} F.,  {Bontemps} S.,
  {Wyrowski} F.,   {et al.} 2014, \aap, 565, 75

\bibitem[\protect\citeauthoryear{{Draine}}{{Draine}}{2003}]{Draine2003}
{Draine} B.~T.,  2003, \araa, 41, 241

\bibitem[\protect\citeauthoryear{{Elia}, {Molinari}, {Schisano}, {Robitaille},
  {Angles-Alcazar}  \& {et al.}}{{Elia} et~al.}{2010}]{elia10}
{Elia} D.,  {Molinari} S.,  {Schisano} E.,  {Robitaille} T.~P.,
  {Angles-Alcazar} D.,   {et al.} 2010, A\&A, 518, L97

\bibitem[\protect\citeauthoryear{{Elia}, {Molinari}, {Schisano}, {Pestalozzi},
  {Merello}  \& {et al.}}{{Elia} et~al.}{2017}]{Elia+2017}
{Elia} D.,  {Molinari} S.,  {Schisano} E.,  {Pestalozzi} M.,  {Merello} M.,
  {et al.} 2017, \mnras, 471, 100

\bibitem[\protect\citeauthoryear{{Elia}, {Molinari}, {Merello}, {Schisano}, {Di
  Giorgio}, {Traficante}, {Liu}  \& {et al.}}{{Elia} et~al.}{2019}]{Elia+2019}
{Elia} D.,  {Molinari} S.,  {Merello} M.,  {Schisano} E.,  {Di Giorgio} A.~M.,
  {Traficante} A.,  {Liu} S.~J.,   {et al.} 2019, in preparation

\bibitem[\protect\citeauthoryear{{Furlan} et~al.,}{{Furlan}
  et~al.}{2016}]{Furlan+2016}
{Furlan} E.,  et~al., 2016, \apjs, 224, 45

\bibitem[\protect\citeauthoryear{{Ginsburg} et~al.,}{{Ginsburg}
  et~al.}{2013}]{Ginsburg2013}
{Ginsburg} A.,  et~al., 2013, \apjs, \href
  {http://adsabs.harvard.edu/abs/2013arXiv1305.6622G} {208, 14}

\bibitem[\protect\citeauthoryear{{Gutermuth}, {Pipher}, {Megeath}, {Myers},
  {Allen}  \& {Allen}}{{Gutermuth} et~al.}{2011}]{Gutermuth2011}
{Gutermuth} R.~A.,  {Pipher} J.~L.,  {Megeath} S.~T.,  {Myers} P.~C.,  {Allen}
  L.~E.,   {Allen} T.~S.,  2011, \mn@doi [\apj] {10.1088/0004-637X/739/2/84},
  \href {http://adsabs.harvard.edu/abs/2011ApJ...739...84G} {739, 84}

\bibitem[\protect\citeauthoryear{{He}, {Zhou}, {Esimbek}, {Ji}, {Wu}, {Tang}
  \& {et al.}}{{He} et~al.}{2015}]{He+2015}
{He} Y.-X.,  {Zhou} J.-J.,  {Esimbek} J.,  {Ji} W.-G.,  {Wu} G.,  {Tang} X.-D.,
    {et al.} 2015, \mnras, 450, 1926

\bibitem[\protect\citeauthoryear{{Heyer}, {Gutermuth}, {Urquhart}, {Csengeri},
  {Wienen}, {Leurini}, {Menten}  \& {Wyrowski}}{{Heyer}
  et~al.}{2016}]{Heyer+2016}
{Heyer} M.,  {Gutermuth} R.~A.,  {Urquhart} J.~S.,  {Csengeri} T.,  {Wienen}
  M.,  {Leurini} S.,  {Menten} K.,   {Wyrowski} F.,  2016, \aap, 588, 29

\bibitem[\protect\citeauthoryear{{Kauffmann}, {Bertoldi}, {Bourke}, {Evans}  \&
  {Lee}}{{Kauffmann} et~al.}{2008}]{Kauffmann+2008}
{Kauffmann} J.,  {Bertoldi} F.,  {Bourke} T.~L.,  {Evans} II N.~J.,   {Lee}
  C.~W.,  2008, \aap, 487, 993

\bibitem[\protect\citeauthoryear{{Kirk}, {Myers}, {Bourke}, {Gutermuth},
  {Hedden}  \& {Wilson}}{{Kirk} et~al.}{2013}]{Kirk+13}
{Kirk} H.,  {Myers} P.,  {Bourke} T.,  {Gutermuth} R.~A.,  {Hedden} A.,
  {Wilson} G.~W.,  2013, \mn@doi [\apj] {10.1088/0004-637X/766/2/115}, \href
  {http://adsabs.harvard.edu/abs/2013ApJ...766..115K} {766, 115}

\bibitem[\protect\citeauthoryear{{Kroupa}}{{Kroupa}}{2001}]{Kroupa2001}
{Kroupa} P.,  2001, \mn@doi [\mnras] {10.1046/j.1365-8711.2001.04022.x}, \href
  {http://adsabs.harvard.edu/abs/2001MNRAS.322..231K} {322, 231}

\bibitem[\protect\citeauthoryear{{McKee} \& {Tan}}{{McKee} \&
  {Tan}}{2003}]{mck03}
{McKee} C.~F.,  {Tan} J.~C.,  2003, \mn@doi [\apj] {10.1086/346149}, \href
  {http://adsabs.harvard.edu/abs/2003ApJ...585..850M} {585, 850}

\bibitem[\protect\citeauthoryear{{Molinari}, {Brand}, {Cesaroni}  \&
  {Palla}}{{Molinari} et~al.}{1996}]{Molinari+1996}
{Molinari} S.,  {Brand} J.,  {Cesaroni} R.,   {Palla} F.,  1996, \aap, 308, 573

\bibitem[\protect\citeauthoryear{{Molinari}, {Testi}, {Rodr{\'{\i}}guez}  \&
  {Zhang}}{{Molinari} et~al.}{2002}]{Molinari+2002}
{Molinari} S.,  {Testi} L.,  {Rodr{\'{\i}}guez} L.~F.,   {Zhang} Q.,  2002,
  \apj, 570, 758

\bibitem[\protect\citeauthoryear{{Molinari}, {Pezzuto}, {Cesaroni}, {Brand},
  {Faustini}  \& {Testi}}{{Molinari} et~al.}{2008}]{Molinari+2008}
{Molinari} S.,  {Pezzuto} S.,  {Cesaroni} R.,  {Brand} J.,  {Faustini} F.,
  {Testi} L.,  2008, \aap, 481, 345

\bibitem[\protect\citeauthoryear{{Molinari} et~al.,}{{Molinari}
  et~al.}{2016a}]{Molinari+2017}
{Molinari} S.,  et~al., 2016a, in {Brescia} M.,  {Djorgovski} G.,  {Feigelson}
  E.~D.,  {Longo} G.,   {Cavuoti} S.,  eds,  Proceedings of IAU Symposium Vol.
  325, Astroinformatics. pp 291--298

\bibitem[\protect\citeauthoryear{{Molinari} et~al.,}{{Molinari}
  et~al.}{2016b}]{Molinari+2016a}
{Molinari} S.,  et~al., 2016b, \aap, 591, 149

\bibitem[\protect\citeauthoryear{{Molinari}, {Merello}, {Elia}, {Cesaroni},
  {Testi}  \& {Robitaille}}{{Molinari} et~al.}{2016c}]{Molinari+2016c}
{Molinari} S.,  {Merello} M.,  {Elia} D.,  {Cesaroni} R.,  {Testi} L.,
  {Robitaille} T.,  2016c, \apjl, 826, L8

\bibitem[\protect\citeauthoryear{{Morales} \& {Robitaille}}{{Morales} \&
  {Robitaille}}{2017}]{Morales+2017}
{Morales} E. F.~E.,  {Robitaille} T.~P.,  2017, \aap, in press

\bibitem[\protect\citeauthoryear{{Palla} \& {Stahler}}{{Palla} \&
  {Stahler}}{1993}]{PS1993}
{Palla} F.,  {Stahler} S.~W.,  1993, \apj, 418, 414

\bibitem[\protect\citeauthoryear{{Parmentier}, {Pfalzner}  \&
  {Grebel}}{{Parmentier} et~al.}{2014}]{Parmentier+2014}
{Parmentier} G.,  {Pfalzner} S.,   {Grebel} E.~K.,  2014, \apj, 791, 132

\bibitem[\protect\citeauthoryear{{Peretto} et~al.,}{{Peretto}
  et~al.}{2013}]{Peretto+2013}
{Peretto} N.,  et~al., 2013, \aap, 555, 112

\bibitem[\protect\citeauthoryear{{Polychroni}, {Schisano}, {Elia}  \& {et
  al.}}{{Polychroni} et~al.}{2013}]{Polychroni+2013}
{Polychroni} D.,  {Schisano} E.,  {Elia} D.,   {et al.} 2013, \apjl, 777, 33

\bibitem[\protect\citeauthoryear{{Purcell} et~al.,}{{Purcell}
  et~al.}{2013}]{Purcell2013}
{Purcell} C.~R.,  et~al., 2013, ApJS, 205, 1

\bibitem[\protect\citeauthoryear{{Robitaille}}{{Robitaille}}{2017}]{Rob2017}
{Robitaille} T.~P.,  2017, \aap, 600, 11

\bibitem[\protect\citeauthoryear{{Robitaille}, {Whitney}, {Indebetouw}, {Wood}
  \& {Denzmore}}{{Robitaille} et~al.}{2006}]{rob06}
{Robitaille} T.~P.,  {Whitney} B.~A.,  {Indebetouw} R.,  {Wood} K.,
  {Denzmore} P.,  2006, \mn@doi [\apjs] {10.1086/508424}, \href
  {http://adsabs.harvard.edu/abs/2006ApJS..167..256R} {167, 256}

\bibitem[\protect\citeauthoryear{{Robitaille}, {Whitney}, {Indebetouw}  \&
  {Wood}}{{Robitaille} et~al.}{2007}]{Robitaille+2007}
{Robitaille} T.~P.,  {Whitney} B.~A.,  {Indebetouw} R.,   {Wood} K.,  2007,
  \apjs, 169, 328

\bibitem[\protect\citeauthoryear{{R\"ollig} et~al.,}{{R\"ollig}
  et~al.}{2011}]{Rollig+2011}
{R\"ollig} M.,  et~al., 2011, \aap, 525, 8

\bibitem[\protect\citeauthoryear{{Schuller} et~al.,}{{Schuller}
  et~al.}{2009}]{Schuller2009}
{Schuller} F.,  et~al., 2009, \mn@doi [\aap] {10.1051/0004-6361/200811568},
  \href {http://adsabs.harvard.edu/abs/2009A%26A...504..415S} {504, 415}

\bibitem[\protect\citeauthoryear{{Tan}, {Beltr\'an}, {Caselli}, {Fontani},
  {Fuente}, {Krumholz}  \& {et al.}}{{Tan} et~al.}{2014}]{Tan+2014}
{Tan} J.~C.,  {Beltr\'an} M.~T.,  {Caselli} P.,  {Fontani} F.,  {Fuente} A.,
  {Krumholz} M.~R.,   {et al.} 2014, in {Beuther} H.,  {Klessen} R.,
  {Dullemond} C.~P.,   {Henning} T.,  eds, Protostars and Planets VI.

\bibitem[\protect\citeauthoryear{{Traficante}, {Fuller}, {Billot},
  {Duarte-Cabral}, {Merello}, {Molinari}, {Peretto}  \&
  {Schisano}}{{Traficante} et~al.}{2017}]{Traficante+2017}
{Traficante} A.,  {Fuller} G.~A.,  {Billot} N.,  {Duarte-Cabral} A.,  {Merello}
  M.,  {Molinari} S.,  {Peretto} N.,   {Schisano} E.,  2017, \mnras, Submitted

\bibitem[\protect\citeauthoryear{{Veneziani} et~al.,}{{Veneziani}
  et~al.}{2013}]{Veneziani+2013}
{Veneziani} M.,  et~al., 2013, \aap, 549, 130

\bibitem[\protect\citeauthoryear{{Veneziani} et~al.,}{{Veneziani}
  et~al.}{2017}]{Veneziani+2017}
{Veneziani} M.,  et~al., 2017, \aap, 599, 7

\bibitem[\protect\citeauthoryear{{Weingartner} \& {Draine}}{{Weingartner} \&
  {Draine}}{2003}]{WD2001}
{Weingartner} J.~C.,  {Draine} B.~T.,  2003, \apj, 548, 296

\bibitem[\protect\citeauthoryear{{Wilcock} et~al.,}{{Wilcock}
  et~al.}{2011}]{Wilcock+2011}
{Wilcock} L.~A.,  et~al., 2011, \aap, 526, 159

\bibitem[\protect\citeauthoryear{{Wright}, {Eisenhardt}, {Mainzer}, {Ressler},
  {Cutri}  \& {et al.}}{{Wright} et~al.}{2010}]{Wright+2010}
{Wright} E.~L.,  {Eisenhardt} P.,  {Mainzer} A.~K.,  {Ressler} M.~E.,  {Cutri}
  R.~M.,   {et al.} 2010, \aj, 140, 1868

\bibitem[\protect\citeauthoryear{{Wu}, {Evans}, {Shirley}  \& {Knez}}{{Wu}
  et~al.}{2010}]{Wu2010}
{Wu} J.,  {Evans} II N.~J.,  {Shirley} Y.~L.,   {Knez} C.,  2010, \mn@doi
  [\apjs] {10.1088/0067-0049/188/2/313}, \href
  {http://adsabs.harvard.edu/abs/2010ApJS..188..313W} {188, 313}

\bibitem[\protect\citeauthoryear{{Yadav}, {Pandey}, {Sharma}, {Ojha}, {Samal},
  {Mallick}  \& {et al.}}{{Yadav} et~al.}{2016}]{Yadav+2016}
{Yadav} R.~K.,  {Pandey} A.~K.,  {Sharma} S.,  {Ojha} D.~K.,  {Samal} R.~M.,
  {Mallick} K.~K.,   {et al.} 2016, \mnras, 461, 2502

\bibitem[\protect\citeauthoryear{{Yun}, {Elia}, {Djupvik}, {Torrelles}  \&
  {Molinari}}{{Yun} et~al.}{2015}]{Yun+2015}
{Yun} J.~L.,  {Elia} D.,  {Djupvik} A.~A.,  {Torrelles} J.~M.,   {Molinari} S.,
   2015, \mnras, 452, 1523

\makeatother
\end{thebibliography}

\appendix

\section{The models grid}

The grid of synthetic SED models is available in FITS format file available at http://vialactea.iaps.inaf.it. Due to its large size, the grid is made available in 10 files containing 2 million models each. the files are named "cluster\_set\_n.fits" (with n=0,9)

Each record contains the model parameters and the correspondent SED precomputed at specific wavelength matched to photometric bands of major ground based and space-borne facilities employed for a variety of large-scale Galactic surveys: Spitzer IRAC and MIPS, WISE, MSX, Herschel PACS and SPIRE, LABOCA\@APEX, Bolocam\@ CSO. A subsequent release will contain the full 1\um -1mm SED. All models fluxes are computed at a fixed distance of 1 kpc and are reported in milliJy.

An additional file named "models\_sed\_ratio.dat" in ASCII format is also made available. It contains 20 million values (1 for each model) of the $L_{\star MIR}$ / [L$_{C, T_D}$-L$_{C, 12K}$] ratio that quantifies the balance between fraction of stellar emitted luminosity that can be processed by intra-clump dust, and the FIR luminosity emitted by clump dust (see \S\ref{balance}). The values are listed in the same order in which one would read all models in sequence starting from "cluster\_set\_0.fits" to "cluster\_set\_9.fits".

The following table lists the fields contained in the models grid.

\onecolumn

\begin{table}
	\centering
	\caption{Description of fields in the models grid.}
	\label{fields}
	\begin{tabular}{lll} 
		\hline
		Column Name & Units & Description\\
		\hline
		Cluster\_ID &  & ID of model run \\
		Clump\_Mass & \msun & Total mass of clump \\
		Compact\_Mass\_Fraction & & Fraction of clump mass locked in compact cores \\
		Compact\_Mass\_Desired & \msun & Total mass in cores (desired) \\
		Compact\_Mass\_Actual & \msun & Total mass in cores (actual) \\
		Clump\_Upper\_Age & x 10$^5$ yrs & Time during which star formation has been active in the clump \\
		Dust\_Temp & K & Dust temperature of clump ISM \\
		Random\_sample & 0-9 & index for independent model realizations at any given primary parameters set (10 each) \\
		Bolometric\_Luminosity & \lsun & Bolometric luminosity of the model \\
		N\_STAR\_TOT & & Total number of simulated cores in the model \\
		N\_STAR\_ZAMS & & Number of ZAMS stars in the present model core realization \\
		M\_STAR\_TOT & \msun & Total stellar mass in model realization \\
		M\_STAR\_ZAMS & \msun & Total mass of ZAMS stars in model realization \\
		L\_STAR\_TOT & \lsun & Total stellar bolometric luminosity in model realization \\
		L\_STAR\_ZAMS & \lsun & Total ZAMS stellar bolometric luminosity in model realization \\
		ZAMS\_MASS\_1 & \msun & Mass of 1$^{st}$ most massive ZAMS star in the model realization \\
		ZAMS\_LUMINOSITY\_1 & \lsun & Luminosity of 1$^{st}$ most massive ZAMS star in the model realization \\
		ZAMS\_TEMPERATURE\_1 & K & Photospheric temperature of 1$^{st}$ most massive ZAMS star in the model realization \\
		ZAMS\_MASS\_2 & \msun & Mass of 2$^{nd}$ most massive ZAMS star in the model realization \\
		ZAMS\_LUMINOSITY\_2 & \lsun & Luminosity of 2$^{nd}$ most massive ZAMS star in the model realization \\
		ZAMS\_TEMPERATURE\_2 & K & Photospheric temperature of 2$^{nd}$ most massive ZAMS star in the model realization \\
		ZAMS\_MASS\_3 & \msun & Mass of 3$^{rd}$ most massive ZAMS star in the model realization \\
		ZAMS\_LUMINOSITY\_3 & \lsun & Luminosity of 3$^{rd}$ most massive ZAMS star in the model realization \\
		ZAMS\_TEMPERATURE\_3 & K & Photospheric temperature of 3$^{rd}$ most massive ZAMS star in the model realization \\
		I1\_STAR\_EXT & mJy &  \parbox[t]{11cm}{Total clump-extincted flux from embedded cores synthetic population in Spitzer/IRAC 3.6\um\ band.} \\
		I1\_DUST & mJy & Total flux from clump dust in Spitzer/IRAC 3.6\um\ band. \\
		I1 & mJy & Total (cores+clump) flux in Spitzer/IRAC 3.6\um\ band.\\
		I2\_STAR\_EXT & mJy & \parbox[t]{11cm}{Total clump-extincted flux from embedded cores synthetic population in Spitzer/IRAC 4.5\um\ band.} \\
		I2\_DUST & mJy & Total flux from clump dust in Spitzer/IRAC 4.5\um\ band. \\
		I2 & mJy & Total (cores+clump) flux in Spitzer/IRAC 4.5\um\ band.\\
		I3\_STAR\_EXT & mJy & \parbox[t]{11cm}{Total clump-extincted flux from embedded cores synthetic population in Spitzer/IRAC 5.8\um\ band.} \\
		I3\_DUST & mJy & Total flux from clump dust in Spitzer/IRAC 5.8\um\ band. \\
		I3 & mJy & Total (cores+clump) flux in Spitzer/IRAC 5.8\um\ band.\\
		I4\_STAR\_EXT & mJy & \parbox[t]{11cm}{Total clump-extincted flux from embedded cores synthetic population in Spitzer/IRAC 8.0\um\ band.} \\
		I4\_DUST & mJy & Total flux from clump dust in Spitzer/IRAC 8.0\um\ band. \\
		I4 & mJy & Total (cores+clump) flux in Spitzer/IRAC 8.0\um\ band.\\
		M1\_STAR\_EXT & mJy & \parbox[t]{11cm}{Total clump-extincted flux from embedded cores synthetic population in Spitzer/MIPS 24\um\ band.} \\
		M1\_DUST & mJy & Total flux from clump dust in Spitzer/MIPS 24\um\ band. \\
		M1 & mJy & Total (cores+clump) flux in Spitzer/MIPS 24\um\ band.\\
		WISE1\_STAR\_EXT & mJy & \parbox[t]{11cm}{Total clump-extincted flux from embedded cores synthetic population in WISE 3.4\um\ band.} \\
		WISE1\_DUST & mJy & Total flux from clump dust in WISE 3.4\um\ band. \\
		WISE1 & mJy & Total (cores+clump) flux in Spitzer/IRAC 3.4\um\ band.\\
		WISE2\_STAR\_EXT & mJy & \parbox[t]{11cm}{Total clump-extincted flux from embedded cores synthetic population in WISE 4.6\um\ band.} \\
		WISE2\_DUST & mJy & Total flux from clump dust in WISE 4.6\um\ band. \\
		WISE2 & mJy & Total (cores+clump) flux in WISE 4.6\um\ band.\\
		WISE3\_STAR\_EXT & mJy & \parbox[t]{11cm}{Total clump-extincted flux from embedded cores synthetic population in WISE 5.8\um\ band.} \\
		WISE3\_DUST & mJy & Total flux from clump dust in WISE 5.8\um\ band. \\
		WISE3 & mJy & Total (cores+clump) flux in WISE 5.8\um\ band.\\
		WISE4\_STAR\_EXT & mJy & \parbox[t]{11cm}{Total clump-extincted flux from embedded cores synthetic population in WISE 12\um\ band.} \\
		WISE4\_DUST & mJy & Total flux from clump dust in WISE 12\um\ band. \\
		WISE4 & mJy & Total (cores+clump) flux in WISE 12\um\ band.\\
		\hline
	\end{tabular}
\end{table}

\begin{table}
	\centering
	\contcaption{Description of fields in the models grid.}
	\label{fields}
	\begin{tabular}{lll} 
		\hline
		Column Name & Units & Description\\
		\hline
		XA\_STAR\_EXT & mJy & \parbox[t]{11cm}{Total clump-extincted flux from embedded cores synthetic population in MSX 8.3\um\ band.} \\
		XA\_DUST & mJy & Total flux from clump dust in MSX 8.3\um\ band. \\
		XA & mJy & Total (cores+clump) flux in MSX 8.3\um\ band.\\
		XC\_STAR\_EXT & mJy & \parbox[t]{11cm}{Total clump-extincted flux from embedded cores synthetic population in MSX 12.1\um\ band.} \\
		XC\_DUST & mJy & Total flux from clump dust in MSX 12.1\um\ band. \\
		XC & mJy & Total (cores+clump) flux in MSX 12.1\um\ band.\\
		XD\_STAR\_EXT & mJy & \parbox[t]{11cm}{Total clump-extincted flux from embedded cores synthetic population in MSX 14.6\um\ band.} \\
		XD\_DUST & mJy & Total flux from clump dust in MSX 14.6\um\ band. \\
		XD & mJy & Total (cores+clump) flux in MSX 14.6\um\ band.\\
		XE\_STAR\_EXT & mJy & \parbox[t]{11cm}{Total clump-extincted flux from embedded cores synthetic population in MSX 21.3\um\ band.} \\
		XE\_DUST & mJy & Total flux from clump dust in MSX 21.3\um\ band. \\
		XE & mJy & Total (cores+clump) flux in MSX 21.3\um\ band.\\
		PACS1\_STAR\_EXT & mJy & \parbox[t]{11cm}{Total clump-extincted flux from embedded cores synthetic population in Herschel/PACS 70\um\ band.} \\
		PACS1\_DUST & mJy & Total flux from clump dust in Herschel/PACS 70\um\ band. \\
		PACS1 & mJy & Total (cores+clump) flux in Herschel/PACS 70\um\ band.\\
		PACS2\_STAR\_EXT & mJy & \parbox[t]{11cm}{Total clump-extincted flux from embedded cores synthetic population in Herschel/PACS 100\um\ band.} \\
		PACS2\_DUST & mJy & Total flux from clump dust in Herschel/PACS 100\um\ band. \\
		PACS2 & mJy & Total (cores+clump) flux in Herschel/PACS 100\um\ band.\\
		PACS3\_STAR\_EXT & mJy & \parbox[t]{11cm}{Total clump-extincted flux from embedded cores synthetic population in Herschel/PACS 160\um\ band.} \\
		PACS3\_DUST & mJy & Total flux from clump dust in Herschel/PACS 160\um\ band. \\
		PACS3 & mJy & Total (cores+clump) flux in Herschel/PACS 160\um\ band.\\
		SPIR1\_STAR\_EXT & mJy & \parbox[t]{11cm}{Total clump-extincted flux from embedded cores synthetic population in Herschel/SPIRE 250\um\ band.} \\
		SPIR1\_DUST & mJy & Total flux from clump dust in Herschel/SPIRE 250\um\ band. \\
		SPIR1 & mJy & Total (cores+clump) flux in Herschel/SPIRE 250\um\ band.\\
		SPIR2\_STAR\_EXT & mJy & \parbox[t]{11cm}{Total clump-extincted flux from embedded cores synthetic population in Herschel/SPIRE 350\um\ band.} \\
		SPIR2\_DUST & mJy & Total flux from clump dust in Herschel/SPIRE 350\um\ band. \\
		SPIR2 & mJy & Total (cores+clump) flux in Herschel/SPIRE 350\um\ band.\\
		SPIR3\_STAR\_EXT & mJy & \parbox[t]{11cm}{Total clump-extincted flux from embedded cores synthetic population in Herschel/SPIRE 500\um\ band.} \\
		SPIR3\_DUST & mJy & Total flux from clump dust in Herschel/SPIRE 500\um\ band. \\
		SPIR3 & mJy & Total (cores+clump) flux in Herschel/SPIRE 500\um\ band.\\
		LABOC\_STAR\_EXT & mJy & \parbox[t]{11cm}{Total clump-extincted flux from embedded cores synthetic population in Apex/LABOCA 870\um\ band.} \\
		LABOC\_DUST & mJy & Total flux from clump dust in Apex/LABOCA 870\um\ band. \\
		LABOC & mJy & Total (cores+clump) flux in Apex/LABOCA 870\um\ band.\\
		BOL11\_STAR\_EXT & mJy & \parbox[t]{11cm}{Total clump-extincted flux from embedded cores synthetic population in CSO/Bolocam 1.1mm band.} \\
		BOL11\_DUST & mJy & Total flux from clump dust in CSO/Bolocam 1.1mm band. \\
		BOL11 & mJy & Total (cores+clump) flux in CSO/Bolocam 1.1mm band.\\		
		\hline
	\end{tabular}
\end{table}

\bsp	
\label{lastpage}
\end{document}